\documentclass[preprint,pre,oneside,floatfix,endfloats*]{revtex4}

\usepackage{amsmath}
\usepackage{amssymb}
\usepackage{graphicx}
\usepackage{dcolumn}
\usepackage{bm}
\usepackage{color}
\usepackage[dvipsnames]{xcolor}
\usepackage[english]{babel}
\newcommand{\dd}{\ensuremath{\mathop{}\!\mathrm{d}}}
\newcommand{\dps}{\displaystyle}
\newcommand{\bs}[1]{\boldsymbol{#1}}
\newcommand{\rt}[1]{\textcolor{newalert}{#1}}

\definecolor{newalert}{RGB}{255, 10, 10}
\definecolor{newtext}{HTML}{0022CA}
\begin{document}


\title{Effective Landau theory of ferronematics}

\author{Grigorii Zarubin}
\email{zarubin@is.mpg.de}
\author{Markus Bier}
\email{bier@is.mpg.de}
\author{S.\ Dietrich}
\email{dietrich@is.mpg.de}
\affiliation
{
   Max-Planck-Institut f\"ur Intelligente Systeme, 
   Heisenbergstr.\ 3,
   70569 Stuttgart,
   \mbox{Germany, and}
   \uppercase\expandafter{\romannumeral 4\relax}.~Institut f\"ur Theoretische Physik,
   Universit\"at Stuttgart,
   Pfaffenwaldring 57,
   70569 Stuttgart,
   Germany
}

\date{\today}

\begin{abstract}
An effective Landau-like description of ferronematics, i.e., suspensions of
magnetic colloidal particles in a nematic liquid crystal (NLC), is developed in terms of the
corresponding magnetization and nematic director fields. The study is based on a
microscopic model and on classical density functional theory.
Ferronematics are susceptible to weak magnetic fields and they can exhibit a
ferromagnetic phase, which has been predicted several decades ago
and which has recently been found experimentally.
Within the proposed effective Landau theory of ferronematics one has 
quantitative access, e.g., to the coupling between the magnetization of the
magnetic colloids and the nematic director of the NLC. On mesoscopic length scales this generates complex
response patterns.
\end{abstract}

\maketitle


\section{Introduction}

The quest for soft matter systems which exhibit spontaneous symmetry breaking
in terms of a polar order parameter, analogous to ferromagnetism in solids, has a
long history.
The first class of systems investigated under this perspective are suspensions
of magnetic nanoparticles in simple liquids, which exhibit particularly rich 
structural and dynamical properties generated by the intricacies of the dipolar
character of their basic mutual interactions \cite{1995_Jonsson, 1997_Groh, 2000_Camp, 2001_Mueller, 2004_Huke, 2004_Mendelev, 2005_Holm, 2008_Rablau, 2008_Trasca, 2015_Szalai}. 
Moreover, they offer a broad range of application prospectives such as in
medicine \cite{2002_Alexiou, 2003_Pankhurst, 2002_Brusentsov} and technology \cite{2005_Scherer, 2015_Yao, 2008_Ravaud, 2010_Kim}.
However, whereas such ferrofluids, i.e., colloidal suspensions of magnetic
particles in isotropic liquids, display fascinating behaviors in the presence of
an external magnetic field, actual systems exhibit only zero net magnetization
once the external field is switched off, i.e., there is no occurrence of spontaneous symmetry
breaking \cite{2007_Odenbach, 2012_LopezLopez, 2008_Dery, 2012_Zakinyan}.

A class of soft matter systems, which are indeed able to exhibit nonzero net 
magnetization even in the absence of an external magnetic field, are 
ferronematics, i.e., magnetic colloidal particles suspended in anisotropic
liquids, such as a nematic liquid crystal (NLC).
Whereas this type of system has been studied theoretically almost half a
century ago \cite{1970_Brochard}, its experimental realization has been achieved only recently
\cite{2013_Mertelj}.
The remarkable property of ferronematics is caused by the broken rotational
symmetry of the solvent which implies that the colloids prefer certain
orientations with respect to the nematic director, thus restricting their
individual magnetic moments to certain directions.

Alternatively, it may be conceivable to suspend colloidal particles with 
an \emph{electric} instead of a \emph{magnetic} dipole moment in an NLC and to
study their properties in external \emph{electric} instead of \emph{magnetic}
fields (see Refs.~\cite{2003_Reznikov,2018_Emdadi} and references therein).
However, in contrast to the case of magnetic colloidal particles and magnetic
fields, strong distortions of the NLC are expected to occur in the electric 
analogue, because colloidal particles with electric dipoles strongly polarize their
liquid crystaline environment \cite{2007_Li}, and the molecules of the NLC are highly 
susceptible to external electric fields, too \cite{book_deGennes}.
Hence it appears advantageous to focus on ferronematics instead of the more
complicated suspensions of colloidal particles with electric dipole moments in
an NLC.

Exploiting the full range of properties of ferronematics requires a reliable
theoretical description which allows one to infer the mesoscopic structures
formed by these colloidal suspensions from microscopic molecular properties of
the liquid crystalline and colloidal materials.
So far, such a formalism has not been established. Accordingly, the goal of the
present work is to introduce a systematic approach to solve this multi-scale
problem for the case of \textit{dilute} suspensions of magnetic colloids.

In order to describe ferronematic phases an expression for the free
energy of the suspension of magnetic anisotropic colloids in an NLC is required. 
The authors of Ref.~\cite{2013_Mertelj} have proposed a phenomenological form
of such a free energy density in terms of the local magnetization field and the
local nematic director field. 
Here, a similar form of the free energy density is derived by
starting, however, from a microscopic model. 
This enables one to relate the corresponding expansion coefficients of the free energy to material
properties of the colloids and of the liquid crystal.
In order to achieve this goal, a microscopic description of the 
interaction between a single colloidal particle and the surrounding liquid is
considered.
As an illustration the focus is on a simplified model of a single circular 
disc-shaped colloidal particle suspended in an NLC. 
Here, the quantity of interest is the free energy as a function of particle 
orientation with respect to the nematic director far away from the colloid. 
The theory is formulated in terms of a dimensionless \textit{coupling constant} $c$, which 
is proportional to the particle size and which is small ($c<0.1$) for the colloids
used in the experiment reported in Ref.~\cite{2013_Mertelj} (platelet radius 
$\approx 35\,\text{nm}$). 
Here, analytical expressions of the perturbations of the nematic director 
profile up to first order and of the corresponding free energy up to second
order in the coupling parameter $c$ are derived (Sec.~\ref{sec:Theory1}
and Appendix~\ref{sec:A1}).
Numerical calculations are used in order to assess the accuracy of the proposed
perturbation expansion.

This microscopic expression for the free energy of a single colloidal particle in an
NLC can be interpreted from the mesoscopic point of view as an external
one-particle potential the NLC medium exerts onto each colloid.
This one-particle potential can be incorporated into a classical density functional
description of a fluid of magnetic discs suspended in the NLC. 
In agreement with the experimental set-up in Ref.~\cite{2013_Mertelj}, the
present work is restricted to the case of dilute colloidal suspensions, which
allows one to neglect the {effective} interactions between two colloidal particles in order
to gain calculational advantages.
The resulting mesoscopic free energy density is a second degree polynomial of
the local magnetization $\mathbf{M}(\mathbf{r})$ and of the local nematic director 
$\mathbf{n}(\mathbf{r})$ (Sec.~\ref{sec:Theory2} and Appendix~\ref{sec:A2})
which can be directly compared with the corresponding form proposed in 
Ref.~\cite{2013_Mertelj}.

The article is organized as follows. 
In Sec.~\ref{sec:Theory} and in the Appendices \ref{sec:A1} and \ref{sec:A2} the mathematical models are 
introduced in order to be able to investigate the effective one-particle potential of a single, arbitrarily 
thin disc immersed in the NLC and to establish a mesoscopic theory of a dilute 
ferronematic. 
In Sec.~\ref{sec:Res} the results of a numerical assessment of the proposed
effective one-particle potential are presented and the free-energy 
functional of a ferronematic as derived here is compared with the one proposed in 
Ref.~\cite{2013_Mertelj}.
Conclusions and final remarks are given in Sec.~\ref{sec:Discus}.


\section{\label{sec:Theory}Theory}

\subsection{\label{sec:Theory0}Noninteracting particles in a nematic liquid crystal}

\begin{figure}[t]
   \includegraphics[scale=1]{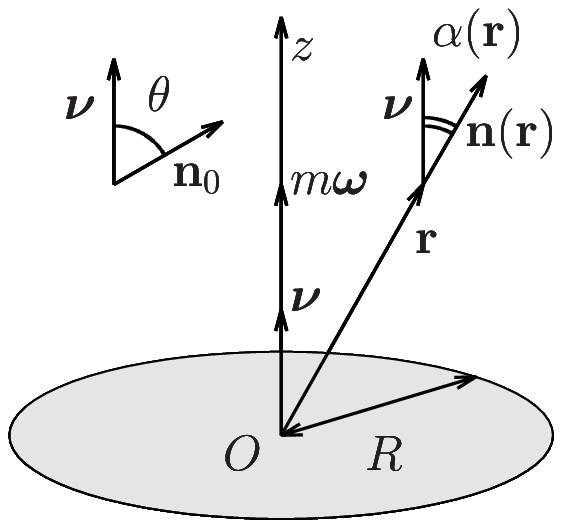}
   \caption{Frame of reference used in the calculation of the free energy of
           a single colloidal disc immersed in an NLC.
           An arbitrarily thin disc of radius $R$ is placed with its center in
           the origin $O$ and is oriented with its normal $\bs{\nu}$ parallel to the $z$-axis, i.e., the
           outer normal to the surface of the disc is denoted as $\bs{\nu}$. 
           A point magnetic dipole of magnitude $m$ and direction $\bs{\omega}||\bs{\nu}$ is placed in the center of the disc.
           The angle between the disc normal $\bs{\nu}$ and the far-field director
           $\mathbf{n}_0$ is denoted as $\theta$, whereas the angle between the disc 
           normal $\bs{\nu}$ and the nematic director $\mathbf{n}(\mathbf{r})$ at any point 
           $\mathbf{r}$ is denoted as $\alpha(\mathbf{r})$.}
   \label{fig:Colloid}
\end{figure}

\begin{figure}[t]
   \includegraphics[scale=0.7]{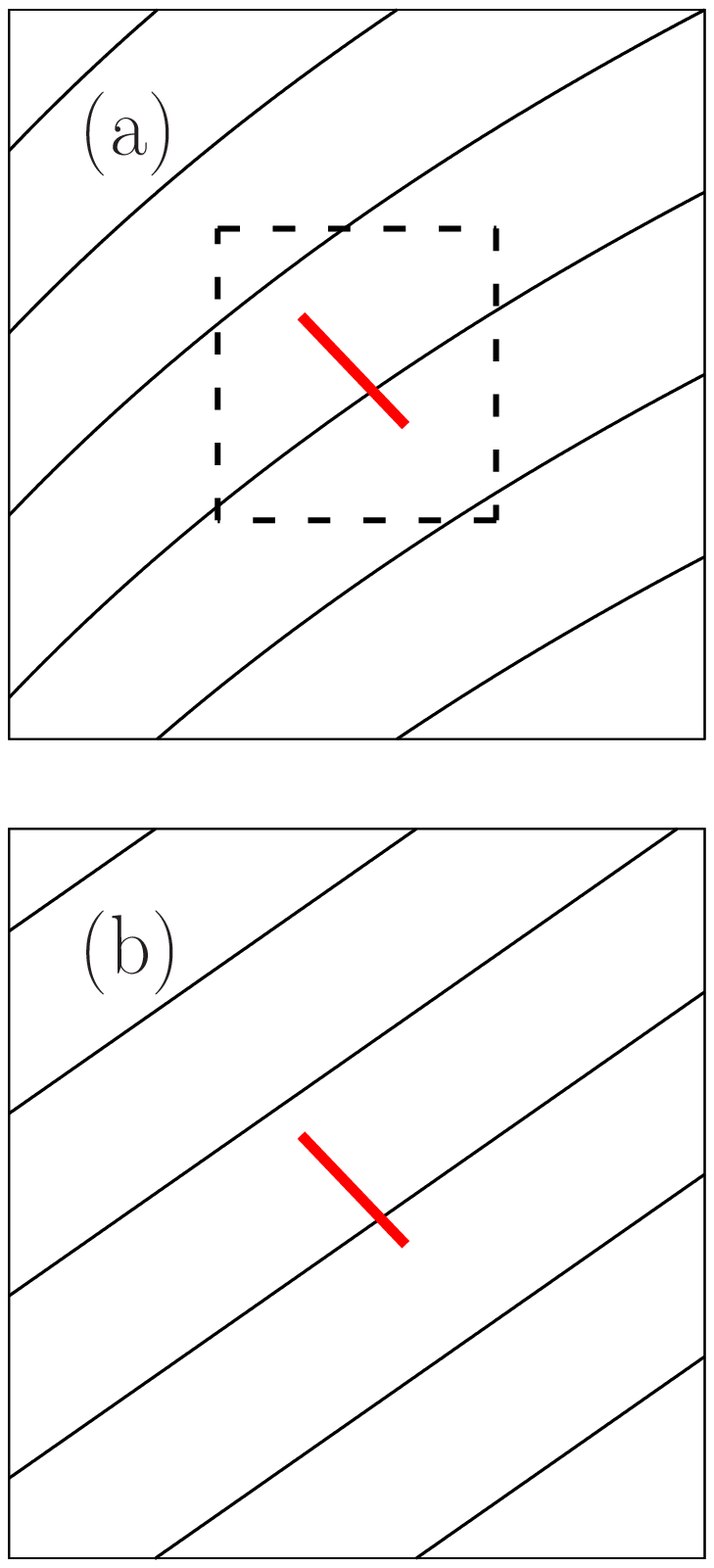}
   \caption{Illustration of the procedure used in the present study in order to model the one-particle elastic potential $\beta V_\text{NLC}(\bs{\omega},\mathbf{n}(\mathbf{r}))$ in Eq.~\eqref{eq:MicroF}. (a): Sideview of a disc-like colloidal particle of radius $R$ at position $\mathbf{r}$ (red rod of length $2R$) is immersed in the NLC with the director field varying slowly on microscopic scales (black solid lines). The region of the slow variation of the director is indicated by the dashed square within which $|\nabla\mathbf{n}(\mathbf{r})|R \ll 1$ is satisfied. (b) The approximately constant field $\mathbf{n}(\mathbf{r})$ (i.e., the black solid lines within the dashed square in (a)) is extrapolated to everywhere and now plays the role of the far-field uniform director field $\mathbf{n}_0$ (black, solid, parallel, and straight lines) used in the complementary problem of a single disc immersed in an NLC (see Subsec.~\ref{sec:Theory1}). The microscopic distortion of the nematic director due to its coupling to the surface of the colloid is not shown.}
   \label{fig:Rational}
\end{figure}

As a first step, we consider a collection of magnetic colloids immersed in an NLC. 
Each colloid is represented by an arbitrarily thin disc of radius $R$ with the outer normal $\bs{\nu}$ to the surface (see Fig.~\ref{fig:Colloid}). A point magnetic dipole of strength $m$ and direction $\bs{\omega}||\bs{\nu}$ is placed in the center of the disc. (Note that the direction of $\bs{\nu}$ depends on which side of the disc is considered whereas the direction of $\bs{\omega}$ does not.) The position of the colloid is the position of its center and the orientation of the colloid is the direction of its magnetic dipole $\bs{\omega}$. (In a more general model $\bs{\omega}$ and $\bs{\nu}$ form a nonzero angle.)

There are four main types of interaction betwen such colloids: the magnetic 
dipole-dipole interaction, the effective interaction induced by the 
elasticity of the NLC medium, steric hard-core interactions, and the van der Waals interaction. 
Here we consider very dilute suspensions of magnetic particles, the volume
fractions $\phi \approx 3 \times 10^{-4}$ of which are comparable with those in 
Ref.~\cite{2013_Mertelj}. 
For such small densities, the dipole-dipole interaction between two colloids
with magnetic moments $m \approx 3 \times 10^{-18}\,\mathrm{Am^2}$ (see 
Ref.~\cite{2013_Mertelj}) and the van der Waals interaction can be neglected 
\cite{footnote1}. Moreover, the steric interaction is disregarded due to its short range and hence the very small impact on the properties of such a dilute solution. 
Here, the effective colloid interaction induced by the NLC elasticity can also be 
neglected due to the high dilution of the suspension and
the weak coupling of colloids to the NLC matrix (see Sec.~\ref{sec:Theory1}).
Therefore, as direct colloid-colloid interactions are negligible for the
type of systems considered here, on the
mesoscopic level the colloidal fluid can be described as an ideal gas in an external field generated by the NLC. The corresponding grand potential functional in
terms of the number density $\rho(\mathbf{r},\bs{\omega})$ of colloids at
position $\mathbf{r}$ and with orientation $\bs{\omega}$ is given by
\begin{align}
   \beta \Omega[\rho,\mathbf{n}] =
   &\ \!\int\!\! \dd^3 r\!\!\int\!\! \dd^2\omega\,\, 
   \rho(\mathbf{r},\bs{\omega})
   \Big[\ln\big(\rho(\mathbf{r},\bs{\omega})\Lambda^3\big) - 1 - \beta\mu
   \notag\\
   &\ + \beta V_\text{NLC}(\bs{\omega},\mathbf{n}(\mathbf{r}))
   - \beta m\bs{\omega} \cdot \mathbf{B}\Big],
   \label{eq:MicroF}
\end{align}
where $\Lambda$ denotes the thermal de Broglie wave length, $\mu$ is the
chemical potential of the colloids, $\mathbf{B}$ describes a uniform external magnetic 
field acting on a magnetic dipole of strength $m$ and orientation $\boldsymbol{\omega}$ and $\beta V_\text{NLC}(\bs{\omega},\mathbf{n}(\mathbf{r}))$ is the one particle external field which describes the coupling of a colloid at position $\mathbf{r}$ and with orientation $\bs{\omega}$ to the NLC with the local director $\mathbf{n}(\mathbf{r})$ at the position of the colloid \cite{footnote2}.

The form of $\beta V_\text{NLC}(\bs{\omega},\mathbf{n}(\mathbf{r}))$ is not known a priori; therefore we adopt certain assumptions in order to model it:
whereas the nematic director field $\mathbf{n}(\mathbf{r})$ may be nonuniform
on mesoscopic length scales, it is assumed to vary slowly on 
the scale of the colloid: $|\nabla\mathbf{n}(\mathbf{r})|R \ll 1$. The colloids are separated far from each other due to their low number density. Thus it is assumed that the interaction of a particular colloid with the NLC is determined by the director field in the close vicinity of the colloid. Accordingly, in order to obtain an expression for the one particle elastic potential $\beta V_\text{NLC}(\bs{\omega},\mathbf{n}(\mathbf{r}))$ (see, c.f., Eq.~\eqref{eq:VNLC}), the particular case of an isolated, disc-like colloid immersed in a uniform director field $\mathbf{n}(\mathbf{r})=\mathbf{n}_0=\text{const}$ (see Fig.~\ref{fig:Rational}) is considered in Sec.~\ref{sec:Theory1} below.
Hence, a colloid with orientation $\bs{\omega}$ placed at position $\mathbf{r}$
experiences a one-particle potential (see, c.f., Eq.~\eqref{eq:VNLC}) which is obtained from, c.f., Eq.~\eqref{eq:FWC} by replacing the
microscopically homogeneous far-field director $\mathbf{n}_0$ with
the mesoscopic local director $\mathbf{n}(\mathbf{r})$. 


\subsection{\label{sec:Theory1}Single disc-like colloid immersed in a nematic liquid crystal}

In this subsection we consider a single disc-like colloidal particle of radius $R$ which is suspended in an NLC described by a mesoscopic director
field $\mathbf{n}(\mathbf{r})$.
According to Fig.~\ref{fig:Colloid}, a frame of reference is attached to the 
colloid such that the $z$-axis is parallel to the  normal of the particle.
Here, we study the case of homeotropic boundary conditions on the surface of the colloid, i.e., it is energetically favorable for the director field 
$\mathbf{n}(\mathbf{r})$ at the surface to be parallel to the normal of the
colloid (i.e., the $z$-axis). 
It is the aim of the present subsection to determine the free energy of the system
as function of the colloid orientation, which is described by
an angle $\theta$ between the $z$-axis and the uniform director $\mathbf{n}_0$
far away from the colloid (see Fig.~\ref{fig:Colloid}).

Due to the local inversion symmetry of the nematic phase (i.e., due to the nematic
directors $\mathbf{n}(\mathbf{r})$ and $-\mathbf{n}(\mathbf{r})$ describing the
same thermodynamic state of the NLC \cite{book_deGennes}), both the angles $\theta=0$ and 
$\theta=\pi$ correspond to the free-energetic ``ground'' state, i.e., the director field
$\mathbf{n}(\mathbf{r})$ at any point $\mathbf{r}$ of the colloid surface points
along its surface normal. Moreover, since there are no deformations in the bulk NLC, the 
director field $\mathbf{n}(\mathbf{r})$ is uniform \textit{everywhere}, and the elastic
free energy of the NLC attains its minimum as a function of $\theta$.
In order to determine the free energy as function of $\theta\in [0,\pi/2]$ (the
behavior in the range $\theta\in [\pi/2,\pi]$ follows from the symmetry of the
free energy with respect to $\theta=\pi/2$ due to the local
inversion symmetry of the NLC), it is obviously necessary to include the
coupling of the director field $\mathbf{n}(\mathbf{r})$ to the particle surface.

In the limit, which is called ``infinite anchoring'', in the following the 
director field $\mathbf{n}(\mathbf{r})$ at the colloid surface is kept fixed to a 
certain (here the normal) direction called ``easy axis'' \cite{book_deGennes}. 
In the language of boundary value problems this limit corresponds to a
Dirichlet boundary condition at the surface of the particle. 
If this constraint is relaxed and the director field $\mathbf{n}(\mathbf{r})$
at the surface can deviate from the easy axis, the free energy acquires an extra
contribution penalizing deviations from the easy direction:
\begin{equation}
   F_{\text{s}} = 
   W\int_{\text{disc}}\!\!\!\!\!\!\dd^2 s\,\, 
   [\mathbf{n}(\mathbf{s})\times\boldsymbol{\nu}(\mathbf{s})]^2,
   \label{eq:FS}
\end{equation}
where $\mathbf{n}(\mathbf{s})$ is the director field at the point $\mathbf{s}$ of the disc surface, $\boldsymbol{\nu}(\mathbf{s})$ is the easy axis which here is taken to be normal to the surface, and 
$W=\text{const}>0$ is the anchoring strength with the dimension energy per surface
area.

In the present context, the infinite anchoring limit has been investigated
before. 
Within the infinite anchoring (IA) limit the free energy has the form 
\cite{1976_Hayes}
\begin{align}
   \frac{F^\text{IA}}{KR} &= 4\theta^2\text{, }\theta\in[0,\pi/2],
   \label{eq:Hayes}
\end{align}
where $K$ denotes the Frank elastic constant of the NLC with dimension energy
per length (i.e., force) within the one-constant approximation 
\cite{book_deGennes}. 
The opposite limit,~i.e., the limit of weak anchoring, has not been investigated systematically in the case of discs. This limit, which we shall refer to as ``weak anchoring``, is the relevant one for the colloids used in the experiment reported in Ref.~\cite{2013_Mertelj} (see below).
It turns out (see below) that it is beneficial to formulate the description of the weak anchoring limit as an expansion of the free energy in terms of the dimensionless
coupling constant
\begin{equation}
  \label{eq:C}
  c:=\frac{WR}{K}.
\end{equation}
In the following, the contributions to the
free energy up to and including the order $\propto c^2$ (see, c.f., Eq.~\eqref{eq:FWC}) are determined.

In order to obtain a systematic expansion of the free energy of the NLC with a
colloidal inclusion in terms of powers of the coupling constant $c$\rt, one can start from
the Frank-Oseen functional of the nematic director field $\mathbf{n}$:
\begin{equation}
  \label{eq:FOFunc}
  \frac{F[\mathbf{n}]}{KR}\!=\! 
  \frac{1}{2R}\int_{\mathcal{V}}\!\!\dd^3 r\,
  \frac{\partial n_i(\mathbf{r})}{\partial x_j}
  \frac{\partial n_i(\mathbf{r})}{\partial x_j} \!+\! 
  \frac{c}{R^2}\int_{\partial\mathcal{V}}\!\!\!\!\dd^2\! s\,
  [\mathbf{n}(\mathbf{s})\times\boldsymbol{\nu}(\mathbf{s})]^2,
\end{equation}
where summation over repeated indices is assumed, $\mathcal{V}\subseteq
\mathbb{R}^3$ is the space filled by the NLC, and $\partial\mathcal{V}$ denotes
the boundary of the NLC (colloid + cell walls). 
If all lengths are measured in units of $R$ (i.e., $\dd^3 r/R^3=
\dd^3\widetilde{r}$, $R\partial/\partial x_j=\partial/\partial
\widetilde{x}_j$, $\dd^2\!s/R^2=\dd^2\widetilde{s}$, $\mathcal{V}/R^3=\widetilde{\mathcal{V}}$ and $\partial\mathcal{V}/R^2=\partial\widetilde{\mathcal{V}}$) the 
dimensionless parameter $c$ is (up to a numerical factor) the ratio of the surface energy (second term on
the right-hand side of Eq.~\eqref{eq:FOFunc}) and the bulk elastic energy 
(first term on the right-hand side of Eq.~\eqref{eq:FOFunc}).
Alternatively, $c$ can be viewed as the ratio of the particle radius $R$ and
the extrapolation length $l:=K/W$ \cite{book_deGennes}. 
Therefore, the coupling constant $c$ measures the cost of free energy for the
director field $\mathbf{n}$ to deviate at the colloid surface from the easy 
axis compared to the cost of free energy for an elastic distortion of the director
field $\mathbf{n}$ in the bulk.
We define the ``weak anchoring'' regime by the condition $c \ll 1$
and note that according to this definition the notion of ``weak'' does not necessarily mean that
the surface anchoring $W$ is small, but rather that the product $WR$ is small
compared to $K$. The latter of which is a material parameter of the particular
NLC, independent of the colloid material or size. 
This implies that for large values of $W$ one can still find $c \ll 1$ for 
sufficiently small particles. 
As a numerical example we consider, in line with Ref.~\cite{2013_Choi}, the realistic range $W\in[0,10^{-4}]\,
\text{N/m}$ of anchoring strengths, the
particle size $R=35\,\text{nm}$, which is roughly the mean of the size 
distribution in Ref.~\cite{2013_Mertelj}, and $K=10^{-11}\,\text{N}$ for the liquid crystal 5CB.
For these material parameters the coupling constants are in the range 
$c\in[0,0.35]$.

Next, one observes in the case of an arbitrarily thin disc, with homeotropic anchoring of arbitrary strength and with normal $\bs{\nu}=\mathbf{e}_z$,
immersed in an NLC with far-field director $\mathbf{n}_0$, that 
the nematic director field $\mathbf{n}(\mathbf{r})$ anywhere inside the
NLC is parallel to the plane spanned by $\mathbf{e}_z$ and $\mathbf{n}_0$ \cite{footnote3},
which, in the following, is, without restriction of generality, taken to be
the $x$-$z$-plane spanned by the unit vectors $\mathbf{e}_x$ and 
$\mathbf{e}_z$.
This allows one to express the nematic director field $\mathbf{n}:\mathcal{V}
\to\mathbb{R}^3$ in terms of a scalar field $\alpha:\mathcal{V}\to\mathbb{R}$
according to
\begin{equation}
  \mathbf{n}(\mathbf{r}) = 
  \mathbf{e}_x\sin(\alpha(\mathbf{r}))+\mathbf{e}_z\cos(\alpha(\mathbf{r})).
  \label{eq:NAnsatz}
\end{equation}
At large distances from the colloid, $|\mathbf{r}|\gg R$, one has the Dirichlet
boundary condition $\alpha(\mathbf{r})\simeq\theta$.
In terms of the scalar field $\alpha$ the free energy functional in
Eq.~\eqref{eq:FOFunc} reads
\begin{equation}
  \label{eq:AFunc}
  \frac{F[\alpha]}{KR} \!=\! 
  \frac{1}{2R}\int_{\mathcal{V}}\!\!\dd^3 r\,
  \left[\boldsymbol{\nabla}\alpha(\mathbf{r})\right]^2 \!+\!
  \frac{c}{R^2}\!\int_{\partial\mathcal{V}}\!\!\!\!\!\dd^2\! s\,
  [\sin(\alpha(\mathbf{s}))]^2.
\end{equation}
The equilibrium state minimizes 
$F[\alpha]/(KR)$ with respect to variations of $\alpha$ which preserve the
Dirichlet boundary condition at large distances. This corresponds to the
Euler-Lagrange equations
\begin{equation}
  \label{eq:BP}
  \begin{cases}
      \boldsymbol{\nabla}^2\alpha(\mathbf{r})=0 
    & \text{, $\mathbf{r}\in\mathcal{V}$}\\
      \dps\boldsymbol{\nabla}\alpha(\mathbf{s})\cdot\boldsymbol{\nu}
      (\mathbf{s})=\frac{c}{R}\sin(2\alpha(\mathbf{s}))
    & \text{, at the disc surfaces}\\
      \alpha(\mathbf{r})\simeq\theta
    & \text{, $|\mathbf{r}|\gg R$}.
  \end{cases}
\end{equation}
The boundary problem posed in Eq.~\eqref{eq:BP} is difficult to solve analytically,
in particular due to the nonlinear expression on the right-hand side of the 
second line in Eq.~\eqref{eq:BP}.
However, for small values of the coupling parameter $c$ it is promising to consider an expansion of
the scalar field $\alpha$ in terms of powers of $c$:
\begin{equation}
  \label{eq:Expan}
  \alpha(\mathbf{r})
  = \sum_{n=0}^{\infty} c^n \alpha^{(n)}(\mathbf{r})
  = \alpha^{(0)}(\mathbf{r}) + c\alpha^{(1)}(\mathbf{r}) + \cdots.
\end{equation}
By inserting the above expansion into Eq.~\eqref{eq:BP} and by comparing
corresponding orders of $c$ one infers boundary problems for $\alpha^{(n)}(\mathbf{r}),
n\in\{0,1,2,\dots\}$.
It turns out that the boundary problems for $\alpha^{(0)}(\mathbf{r})$ and $\alpha^{(1)}(\mathbf{r})$
can be solved analytically (see Appendix \ref{sec:A1}). Accordingly, here we restrict the following
discussion to these two terms of the expansion in
Eq.~\eqref{eq:Expan}. 
Inserting $\alpha^{(0)}(\mathbf{r})$ (Eq.~\eqref{eq:A0}) and $\alpha^{(1)}(\mathbf{r})$ (Eq.~\eqref{eq:A1}) into Eq.~\eqref{eq:AFunc} leads
to the weak anchoring (WA) limit of the free energy (Eq.~\eqref{eq:OneParticleFromA}):
\begin{equation}
  \label{eq:FWC}
  \frac{F^\text{WA}}{KR} =
  -\left(2\pi c+\frac{32}{3}c^2\right)(\mathbf{n}_0\cdot\bs{\omega})^2
  +\frac{32}{3}c^2(\mathbf{n}_0\cdot\bs{\omega})^4.
\end{equation}
It is worth noting that the term $\propto c^1$ in Eq.~\eqref{eq:FWC} can be written in the form
\begin{equation}
-2\pi c(\mathbf{n}_0\cdot\bs{\omega})^2=-2\pi c[\cos\theta]^2=\text{const}+2\pi\frac{WR}{K}[\sin\theta]^2,
\end{equation}
which is equivalent to the expression obtained in Ref.~\cite{2014_Tasinkevych} for the case of a thin rod with tangential anchoring. This fact is related to the topological similarity between the arbitrarily thin disc with homeotropic anchoring and the arbitrarily thin rod with planar anchoring.

In Sec.~\ref{sec:Res} an interval $c\in [0,c_\text{weak}]$ with $c_\text{weak}
\approx 0.1$ is determined numerically such that within this interval Eq.~\eqref{eq:FWC} is a
quantitatively reliable approximation of the exact free energy $F/(KR)$.


\subsection{\label{sec:Theory2}Mesoscopic functional}

\begin{figure}[t!]
   \includegraphics[scale=0.29]{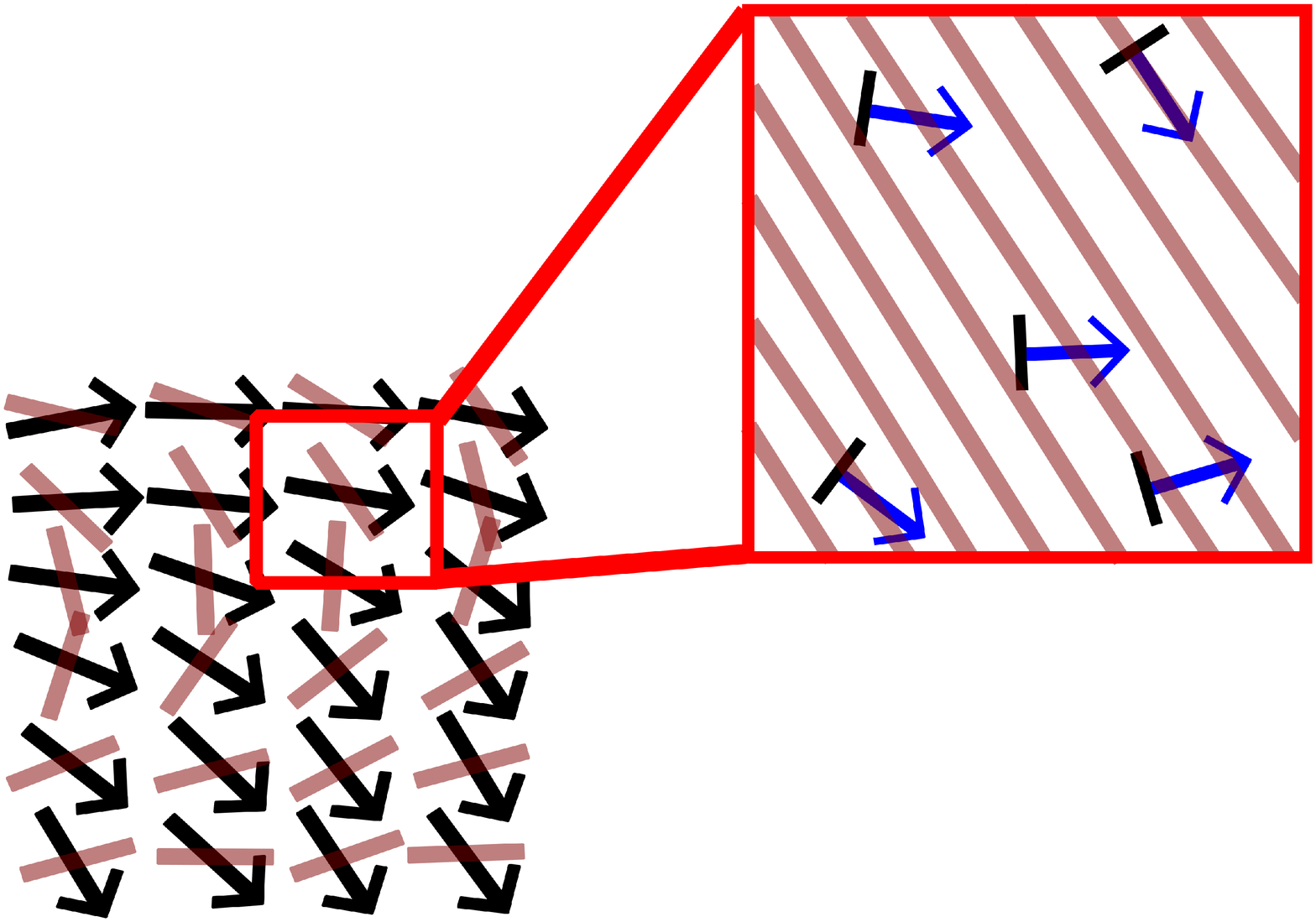}
   \caption{Illustration of the coarse-graining used to merge the microscopic
           and the mesoscopic theory.
           The mesoscopic theory is based on the magnetization field $\mathbf{M}(\mathbf{r})$
           (black arrows in the lower left part) and on the director field $\mathbf{n}(\mathbf{r})$ (pale
           red rods in the lower left part). 
           The mesoscopic magnetization at the position of a volume element
           (red square in the lower left part) is obtained from the magnetic
           moments of individual colloids (blue arrows normal to the black projected discs in the
           upper right part) averaged over the microscopic distribution inside
           the volume element (red square in the upper right part). 
           Within the mesoscopic picture the local director field corresponds to the far-field director $\mathbf{n}_0$ within the microscopic 
           one-particle picture of individual colloids (pale red hatching, see 
           Sec.~\ref{sec:Theory1}). The microscopic distortion of the nematic director due to its coupling to the surface of the colloid is not shown.}
   \label{fig:Model}
\end{figure}

In order to enable a comparison with Ref.~\cite{2013_Mertelj} we aim for replacing the functional $\Omega[\rho,\mathbf{n}]$ of the number density profile
$\rho$ and the nematic director profile $\mathbf{n}$ in Eq.~\eqref{eq:MicroF}
by a functional $\bar{\Omega}[\mathbf{M},\mathbf{n}]:=\Omega[\bar{\rho}(\mathbf{r},\bs{\omega},[\mathbf{M},\mathbf{n}]),\mathbf{n}]$ of the magnetization 
field $\mathbf{M}$ and the nematic director profile $\mathbf{n}$
where $\bar{\rho}(\mathbf{r},\bs{\omega},[\mathbf{M},\mathbf{n}])$ is the 
number density profile of colloids at position $\mathbf{r}$ with orientation 
$\bs{\omega}$ for a prescribed magnetization field $\mathbf{M}(\mathbf{r})$ and
a nematic director field $\mathbf{n}(\mathbf{r})$; $\bar{\rho}$ minimizes the functional in Eq.~\eqref{eq:MicroF}, 
i.e., it is a solution of the Euler-Lagrange equation
\begin{align}
   m(\bs{\omega}\cdot\bs{\lambda}(\mathbf{r},[\mathbf{M},\mathbf{n}]))
   &= 
   \frac{\delta\ \beta\Omega}{\delta \rho(\mathbf{r},\bs{\omega})}
   [\bar{\rho}(\mathbf{r},\bs{\omega},[\mathbf{M},\mathbf{n}]),\mathbf{n}]
   \label{eq:ELEq}\\
   &= \ln(\bar{\rho}(\mathbf{r},\bs{\omega},[\mathbf{M},\mathbf{n}])\Lambda^3)
      - \beta\mu
   \notag\\
   &\phantom{=}\ + \beta V(\bs{\omega},\mathbf{n}(\mathbf{r}),\mathbf{B}),
   \notag
\end{align}
where $\beta V(\bs{\omega},\mathbf{n}(\mathbf{r}),\mathbf{B}):=
\beta V_\text{NLC}(\bs{\omega},\mathbf{n}(\mathbf{r})) - \beta m 
\bs{\omega}\cdot \mathbf{B}$ and where $\bs{\lambda}(\mathbf{r},[\mathbf{M},
\mathbf{n}])$ are the Lagrange multipliers which implement the constraint 
\begin{equation}
  \label{eq:MDef}
  \mathbf{M}(\mathbf{r}) =
  \int\!\!\dd^2\omega\,\, 
  m\bs{\omega} \bar\rho(\mathbf{r},\bs{\omega},[\mathbf{M},\mathbf{n}]).
\end{equation}
According to Subsecs.~\ref{sec:Theory0} and \ref{sec:Theory1}, the external field which the NLC exerts on the fluid of colloidal discs, is described by
\begin{align}
  &\ \beta V_\text{NLC}(\bs{\omega},\mathbf{n}(\mathbf{r}))
  = -(a+b)(\bs{\omega} \cdot \mathbf{n}(\mathbf{r}))^2 +
     b(\bs{\omega} \cdot \mathbf{n}(\mathbf{r}))^4,
  \notag\\
  &\ a:=2\pi\beta KRc, \quad
     b:=\frac{32}{3}\beta KRc^2.
  \label{eq:VNLC}
\end{align}
The solution of Eq.~\eqref{eq:ELEq} is given by
\begin{align}
  &\phantom{=}\ \ \bar{\rho}(\mathbf{r},\bs{\omega},[\mathbf{M},\mathbf{n}])
  \label{eq:rhobar}\\
  &= \zeta\exp\Big(-\beta V(\bs{\omega},\mathbf{n}(\mathbf{r}),\mathbf{B})
  +m\bs{\omega}\cdot\bs{\lambda}(\mathbf{r},[\mathbf{M},\mathbf{n}])\Big)
  \notag
\end{align}
with the fugacity $\zeta:=\exp(\beta\mu)/\Lambda^3$.
Upon inserting Eq.~\eqref{eq:rhobar} into Eq.~\eqref{eq:MicroF} one obtains
\begin{align}
  \beta \Omega[\bar{\rho}(\mathbf{r},\bs{\omega},[\mathbf{M},\mathbf{n}]),\mathbf{n}] =
  &\ \!\int\!\! \dd^3 r\!\!\int\!\! \dd^2\omega\,\, 
  \bar{\rho}(\mathbf{r},\bs{\omega},[\mathbf{M},\mathbf{n}])
  \Big[\beta\mu-\beta V(\bs{\omega},\mathbf{n}(\mathbf{r}),\mathbf{B})
  \notag\\
  &\ +m\bs{\omega}\cdot\bs{\lambda}(\mathbf{r},[\mathbf{M},\mathbf{n}]) - 1 - \beta\mu + \beta V(\bs{\omega},\mathbf{n}(\mathbf{r}),\mathbf{B})\Big]
  \notag\\
  &\ =\int\!\! \dd^3 r\,\,\Big[\bs{\lambda}(\mathbf{r},[\mathbf{M},\mathbf{n}])
  \cdot\int\!\! \dd^2\omega\,\,m\bs{\omega}\bar{\rho}(\mathbf{r},\bs{\omega},[\mathbf{M},\mathbf{n}])
  \notag\\
  &\ -\int\!\! \dd^2\omega\,\,\bar{\rho}(\mathbf{r},\bs{\omega},[\mathbf{M},\mathbf{n}])\Big],
\end{align}
from which it follows that (see Eq.~\eqref{eq:MDef})
\begin{equation}
  \label{eq:NewMicroF}
  \beta \bar{\Omega}[\mathbf{M},\mathbf{n}]=
  \int\!\! \dd^3 r\,\,\Big[\mathbf{M}(\mathbf{r})\cdot
  \bs{\lambda}(\mathbf{r},[\mathbf{M},\mathbf{n}])
  -\rho_0(\mathbf{r},[\mathbf{M},\mathbf{n}])\Big],
\end{equation}
where $\rho_0[\mathbf{M},\mathbf{n}]$ is the orientation-independent number
density profile of the discs:
\begin{equation}
  \label{eq:Rho0}
  \rho_0(\mathbf{r},[\mathbf{M},\mathbf{n}]):=
  \int\!\! \dd^2\omega\,\, 
  \bar{\rho}(\mathbf{r},\bs{\omega},[\mathbf{M},\mathbf{n}]).
\end{equation}
Equation \eqref{eq:MDef} provides the important link which allows one the formulation of the functional of the density profile $\bar\rho(\mathbf{r},\bs{\omega},[\mathbf{M},\mathbf{n}])$ in terms of the mesoscopic magnetization field $\mathbf{M}(\mathbf{r})$. The illustration of this idea is shown in Fig.~\ref{fig:Model}.

In order to derive an explicit expression for $\beta\bar{\Omega}[\mathbf{M},
\mathbf{n}]$ it is convenient to introduce an effective magnetic field
\begin{equation}
  \mathbf{h}(\mathbf{r},[\mathbf{M},\mathbf{n}]):=
  \bs{\lambda}(\mathbf{r},[\mathbf{M},\mathbf{n}])+\beta\mathbf{B}
  \label{eq:EffMF}
\end{equation}
and the generating function
\begin{align}
  &Z(\mathbf{h}):=\int\!\!\dd^2\omega\,\,
  \exp\bigg(A_1(\bs{\omega}\cdot \mathbf{n}(\mathbf{r}))^2 +
            A_2(\bs{\omega}\cdot \mathbf{n}(\mathbf{r}))^4 
  \notag\\
  &\phantom{Z(\mathbf{h}):=\int\!\!\dd^2\omega\,\,\exp\bigg(}
  + m\mathbf{h} \cdot \bs{\omega}\bigg)
  \label{eq:ExactZ}
\end{align}
with $A_1:=a+b$ and $A_2:=-b$ denoting the coefficients of two powers of $(\bs{\omega}\cdot \mathbf{n}(\mathbf{r}))$. 
This leads to (see Eqs.~\eqref{eq:rhobar} and \eqref{eq:Rho0})
\begin{equation}
   \rho_0(\mathbf{r},[\mathbf{M},\mathbf{n}]) 
   = \zeta Z(\mathbf{h}(\mathbf{r},[\mathbf{M},\mathbf{n}]))
   \label{eq:rho0Z}
\end{equation}
and
\begin{equation}
   \mathbf{M}(\mathbf{r})
   = \zeta \frac{\partial Z}{\partial\mathbf{h}}
   (\mathbf{h}(\mathbf{r},[\mathbf{M},\mathbf{n}])).
   \label{eq:MZ}
\end{equation}
With this notation Eq.~\eqref{eq:NewMicroF} reads
\begin{align}
  \beta \bar{\Omega}[\mathbf{M},\mathbf{n}]
  \!=\!\!\int\!\!\dd^3 r\,\bigg[ 
  &\zeta
  \left(\mathbf{h}(\mathbf{r})\cdot
  \frac{\partial Z}{\partial\mathbf{h}}(\mathbf{h}(\mathbf{r})) -
  Z(\mathbf{h}(\mathbf{r}))\right)
  \notag\\
  & - \left.\beta\mathbf{M}(\mathbf{r})\cdot\mathbf{B}\bigg]
  \right|_{\mathbf{h}=\mathbf{h}(\mathbf{r}, [\mathbf{M},\mathbf{n}])}.
  \label{eq:NewNewMicroF}
\end{align}

In the experiments described in Ref.~\cite{2013_Mertelj} the sample is prepared
by dispersing a number density $\rho_\text{iso}$ of colloids in the isotropic
high-temperture phase of the solvent, followed by a quench of the solvent into
the low-temperature nematic phase.
In the absence of an external magnetic field ($\mathbf{B}=0$), the 
magnetization vanishes ($\mathbf{M}=0$) before and after the quench, which
corresponds to the effective magnetic field $\mathbf{h}=0$ \cite{footnote4}.
Noting that the number density of colloids does not change during the quench,
one obtains from Eq.~\eqref{eq:rho0Z}
\begin{equation}
   \zeta = \frac{\rho_0}{Z(0)} = \frac{\rho_\text{iso}}{Y_{00}},
   \label{eq:Fugacity}
\end{equation}
where $Y_{00}$ is defined in Eq.~\eqref{eq:Y00}.

It turns out (see Appendix~\ref{sec:A2}) that this part of the 
integrand in Eq.~\eqref{eq:NewNewMicroF}, which depends on $Z$, is an even function of 
both $H:=m|\mathbf{h}|$ and $u:=\mathbf{n}\cdot\mathbf{h}/|\mathbf{h}|$ (Eq.~\eqref{eq:EffMF}), i.e., a function of $m^2|\mathbf{h}|^2$ and $(\mathbf{n}\cdot\mathbf{h}/|\mathbf{h}|)^2$.
Moreover, it can be shown (see Appendix~\ref{sec:A2}) that the quantities 
$T:=|\mathbf{M}|/(m\zeta)$ and $t:=\mathbf{n}\cdot\mathbf{M}/(m\zeta)$ are both
functions of $H$ and $u$, too. 
If one can invert the map $(H,u)\mapsto(T,t)$, the integrand in 
Eq.~\eqref{eq:NewNewMicroF}, which equals the grand potential density, can be
expressed as function of $|\mathbf{M}|^{2}$ and $(\mathbf{M}\cdot\mathbf{n})^{2}$. 
However, in general inverting this map is very challenging. 
Therefore, the following considerations are restricted to a quadratic 
approximation which includes only terms up to $\sim H^2$ in $Z(\mathbf{h})=: \bar{Z}(H,u)$ (see Eq.~\eqref{eq:quadraticZ}):
\begin{equation}
  \label{eq:ApproxZ}
  \bar{Z}(H,u) \simeq Y_{00} + Y_{10}H^2 + Y_{12}H^2u^2.
\end{equation}
(Note the absence of a term $\propto u^2$.) From Eq.~\eqref{eq:Tandt} in Appendix \ref{sec:A2} one obtains the following system
of equations (see Eq.~\eqref{eq:tTHu}):
\begin{equation}
  \begin{cases}
    \big(t(H,u)\big)^2&=4(Y_{10}+Y_{12})^2 H^2 u^2\\
    \big(T(H,u)\big)^2&=4(2Y_{10}Y_{12}+Y_{12}^2)H^2 u^2+4Y_{10}^2 H^2,
  \end{cases}
  \label{eq:tT}
\end{equation}
which readily can be inverted. 
This renders the grand potential functional within the quadratic approximation:
\begin{align}
  \beta \bar{\Omega}[\mathbf{M},\mathbf{n}]
  &=\int\!\!\dd^3 r\,\,\Big[ 
  \zeta(C_{00} + C_{20}T(\mathbf{r})^2 + C_{02}t(\mathbf{r})^2)
  \notag\\
  &\phantom{= \int\!\!\dd^3 r\,\,\Big[}
  -\beta\mathbf{M}(\mathbf{r})\cdot\mathbf{B}\Big]
  \label{eq:OmegaTt}
\end{align}
with (see Eqs.~\eqref{eq:C00}-\eqref{eq:C02})
\begin{align}
  C_{00}=-Y_{00},\  
  &C_{20}=\frac{1}{4Y_{10}},\ 
  C_{02}=-\frac{Y_{12}/Y_{10}}{4(Y_{10}+Y_{12})},
\end{align}
or, if written explicitly in terms of $|\mathbf{M}|$ and 
$\mathbf{M}\cdot\mathbf{n}$:
\begin{widetext}
  \begin{equation}
    \label{eq:FinalF}
    \beta\bar{\Omega}[\mathbf{M},\mathbf{n}]
    =\int\!\!\dd^3 r\,\,\Big[\rho_\text{iso}\left(
    \frac{C_{00}}{Y_{00}}+ 
    Y_{00}C_{20}\left|\frac{\mathbf{M}(\mathbf{r})}{m\rho_\text{iso}}\right|^2+
    Y_{00}C_{02}\left(\frac{\mathbf{M}(\mathbf{r})}{m\rho_\text{iso}}\cdot
    \mathbf{n}(\mathbf{r})\right)^2\right)
    -\beta\mathbf{M}(\mathbf{r})\cdot\mathbf{B}\Big].
  \end{equation}
\end{widetext}


\section{\label{sec:Res}Results}

\subsection{Limits of reliability for using the one-particle potential}

In the present study we use the expression in Eq.~\eqref{eq:VNLC} for the
one-particle potential $V_\text{NLC}$ corresponding to the weak anchoring regime described by the energy $F^\text{WA}/(KR)$ in Eq.~\eqref{eq:FWC} of a single colloidal particle
with orientation $\bs{\omega}$ at position $\mathbf{r}$, which is immersed in the NLC with
the nematic director $\mathbf{n}(\mathbf{r})$.
In order to assess the accuracy of the expression in Eqs.~\eqref{eq:FWC} or \eqref{eq:VNLC} as 
function of the coupling constant $c$, for comparison the full expression for the free energy 
in Eq.~\eqref{eq:AFunc} is minimized numerically by using a Galerkin finite element
method \cite{1985_Wait}, because analytical solutions of the boundary value
problems for $\alpha^{(j)}, j\geq 2$, are not available.
The specific set-up, which is considered, consists of a cubic box of dimension $30R \times 
30R \times 30R$ which contains a single arbitrarily thin disc in its center. 
The interior of the box is decomposed into tetrahedra and the boundary due to the disc
is decomposed into triangles. 
Within each finite element the unknown function $\alpha(\mathbf{r})$ is
approximated by linear functions which interpolate between its values at the
corners, i.e., the vertices of the triangulation.
Within the finite-dimensional subspace of functions, which are piecewise linear with
respect to the given triangulation, both the volume and the surface integral in
Eq.~\eqref{eq:AFunc} can be calculated explicitly for each finite element. This
allows for a numerical minimization within this finite-dimensional 
subspace. 
The described Galerkin method can be performed for arbitrary values of the
coupling constant $c$, i.e., one is not restricted to the weak anchoring 
regime. 

\begin{figure}[t!]
   \includegraphics[scale=1.0]{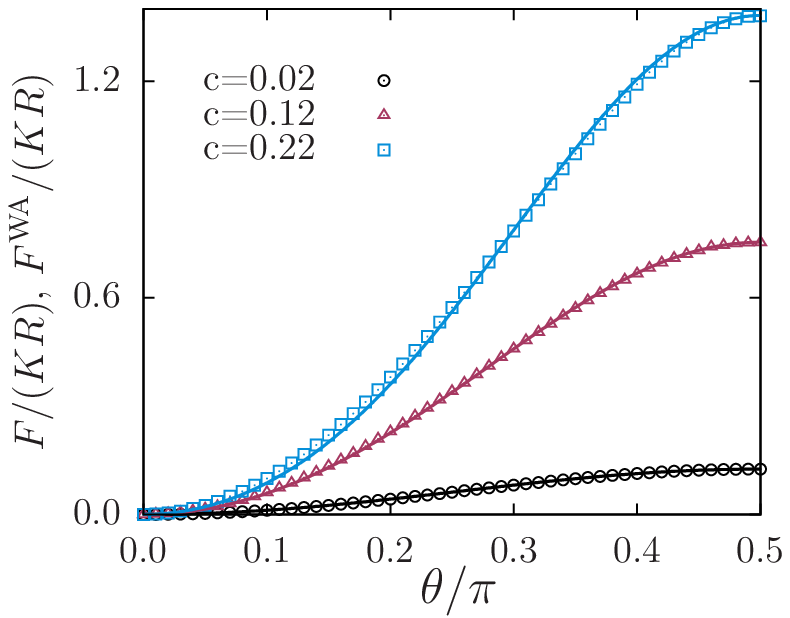}
   \caption{Comparison of the free energy $F/(KR)$ obtained by numerical minimization of 
           Eq.~\eqref{eq:AFunc} (symbols) and the corresponding analytical result $F^\text{WA}/(KR)$ (Eq.~\eqref{eq:FWC}, solid lines) for three values of the 
           coupling constant $c$. 
           Already for $c=0.22$ (blue, squares) the numerical data noticeably
           deviate from the analytic result.}
   \label{fig:Weak}
\end{figure}

Figure \ref{fig:Weak} compares the numerically obtained free energy (symbols)
with the one obtained within the weak anchoring limit (Eq.~\eqref{eq:FWC}, 
solid lines) as function of $\theta$ for three values $c\in\{0.02, 0.12, 
0.22\}$. 
For very small coupling constants (see $c=0.02$) the weak anchoring limit 
Eq.~\eqref{eq:FWC} agrees very well with the numerical results, whereas there are small but
visible deviations for larger values of $c$ (see Fig.~\ref{fig:Weak}, 
$c=0.22$). 

\begin{figure}[t!]
   \centering
   \includegraphics[scale=1.0]{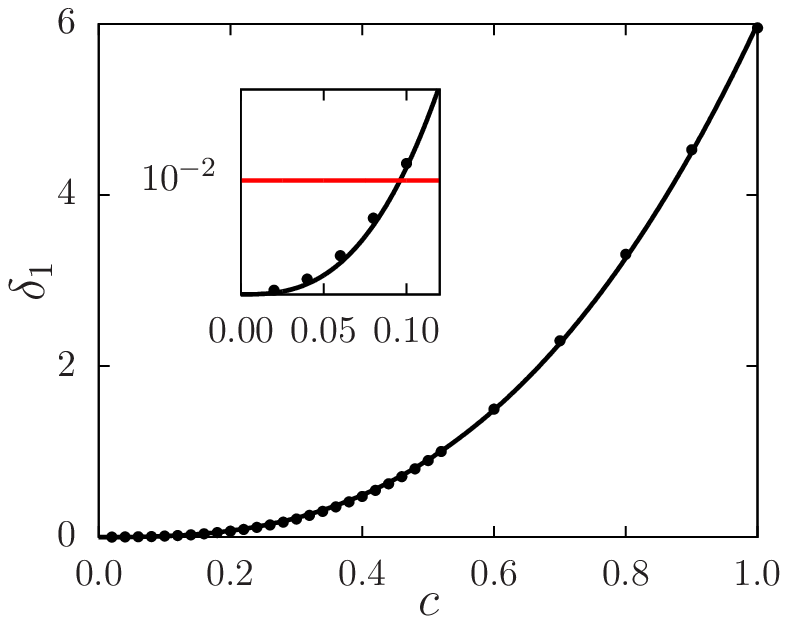}
   \caption{Quadratic norm $\delta_1(c):=\sqrt{\int_0^{\pi/2}d\theta|
           \frac{F^\text{WA}}{KR}-\frac{F}{KR}|^2}$ as function of the coupling
           strength $c$, where $F$ represents the numerically exact free energy obtained from 
           Eq.~\eqref{eq:AFunc} and $F^\text{WA}$ its weak anchoring limit 
           (Eq.~\eqref{eq:FWC}).
           The black solid line is a fit of the power law $\alpha c^{\zeta}$ to the data; $\alpha=6.00\pm 0.01$ and $\zeta=2.73\pm 0.01$.
           The inset is a close-up of the range $c\in [0,0.1]$ illustrating
           the criterion in Eq.~\eqref{eq:Criterion1} for $\varepsilon=10^{-2}$.}
   \label{fig:Criterion1}
\end{figure}

In order to quantify the deviation of the weak anchoring approximation
$F^\text{WA}/(KR)$ in Eq.~\eqref{eq:FWC} from the exact expression $F/(KR)$ in
Eq.~\eqref{eq:AFunc}, the following criterion is introduced 
\cite{footnote5}: 
For a given $\varepsilon>0$ the weak anchoring approximation $F \approx F^\text{WA}$
is considered to be sufficient to describe the free energy $F$ for a fixed value
of $c$, if $\varepsilon$ is an upper bound of the quadratic norm
\begin{equation}
  \label{eq:Criterion1}
  \delta_1(c) :=
  \sqrt{\int_0^{\pi/2}\dd\theta\left|\frac{F^\text{WA}}{KR}-
                                   \frac{F}{KR}\right|^2}<\varepsilon.
\end{equation}
The particular choice of $\varepsilon$ is somewhat arbitrary. 
Figure \ref{fig:Criterion1} shows $\delta_1$ as a function of the coupling
strength $c$. 
The resulting curve can be fitted by a power law $\alpha c^{\zeta}$ with the amplitude 
$\alpha=6.00 \pm 0.01$ and the exponent $\zeta=2.73 \pm 0.01$. 
This fit function allows one to determine a value $c_{\text{weak}}$ such
that the criterion in Eq.~\eqref{eq:Criterion1} with a given tolerance $\varepsilon$
is fulfilled for $c<c_{\text{weak}}$:
\begin{equation}
  c_{\text{weak}} = 
  \left(\frac{\varepsilon}{\alpha}\right)^{1/\zeta} \approx 
  \left(\frac{\varepsilon}{6}\right)^{0.366}.
\end{equation}
A value of, e.g., $\varepsilon=10^{-2}$ implies $c_{\text{weak}} \approx 0.1$
(see the inset in Fig.~\ref{fig:Criterion1}).

Considering Eq.~\eqref{eq:OneParticleFromA}, the contributions to which, up to quadratic order in
$c$, are given by Eq.~\eqref{eq:FWC}, it is tempting to speculate
that the term of cubic order in $c$ is of the form $\sim c^3[\sin(3\theta)]^2$.
In order to assess this presumption one can use the fact that the free energy 
$F(\theta)/(KR)$ in Eq.~\eqref{eq:AFunc} is an even function of $\theta$ with
period $\pi$, which allows for an expansion into a Fourier series
\begin{equation}
  \label{eq:FSeries}
  \frac{F(\theta)}{KR} = \frac{a_0}{2}+\sum_{n=1}^\infty a_n \cos(2n\theta),
\end{equation}
with the Fourier coefficients
\begin{equation}
  \label{eq:FCoeff}
  a_n = 
  \frac{4}{\pi} \int_0^{\pi/2}\!\!\dd \theta\,\, 
  \frac{F(\theta)}{KR}\cos(2n\theta), \quad n=0,1,2,\dots\ .
\end{equation}
In the context of actual numerical schemes only a finite number $N$ of free energy values
$F(\theta_i)/(KR)$ for the angles $\theta_i$, $i\in\{0,1,\dots,N-1\}$, are 
available.
Hence, instead of using Eq.~\eqref{eq:FCoeff} by applying a suitable quadrature,
one can --- as an alternative approximation scheme --- restrict the sum in Eq.~\eqref{eq:FSeries} to $n \leq 
N-1$ and determine the coefficients $a_n$, $n\in\{0,1,\dots,N-1\}$, via fitting the
numerical data $F(\theta_i)/(KR)$, $i\in\{0,1,\dots,N-1\}$, by trigonometric 
polynomials, i.e., superpositions of terms $\cos(2n\theta)$, $n\in\{0,1,\dots,
N-1\}$. 

\begin{figure}[t!]
   \centering
   \includegraphics[scale=1.0]{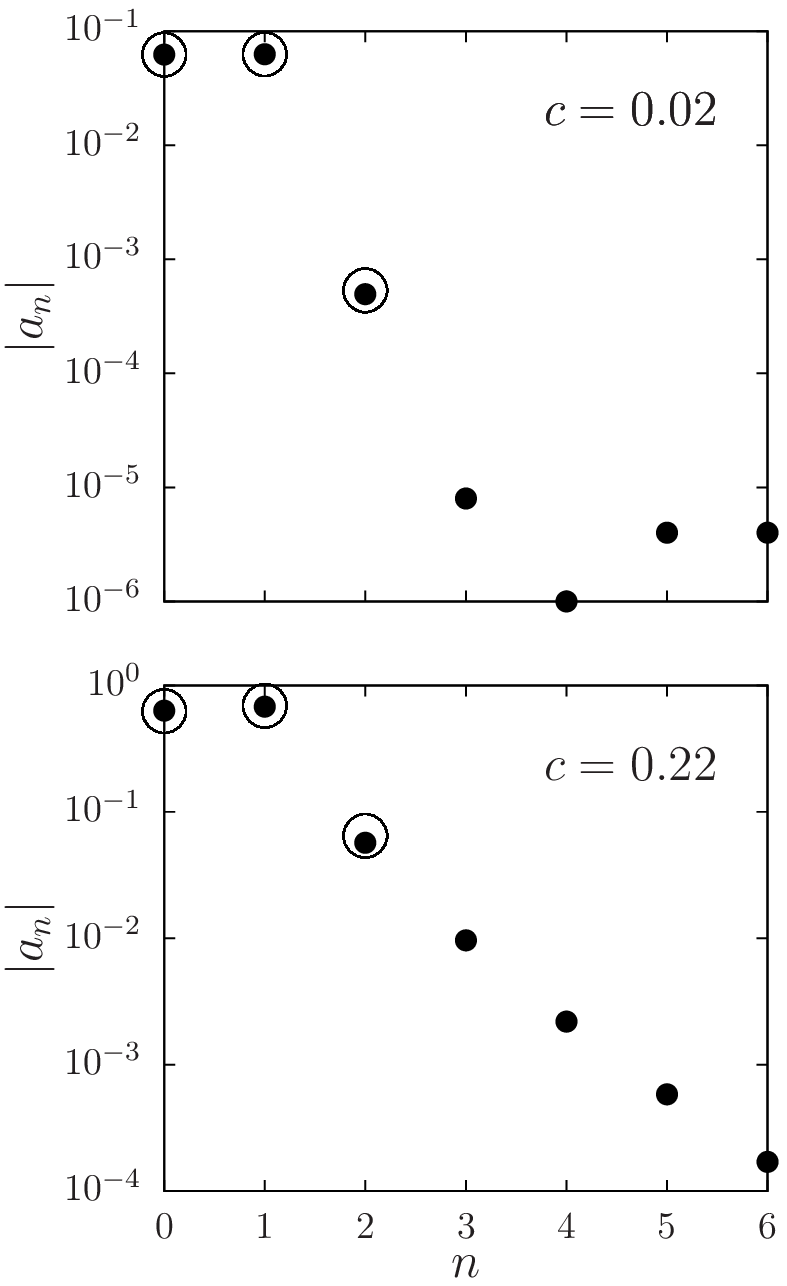}
   \caption{Absolute values $|a_n|$ of the first seven Fourier coefficients of
           the expansion of the exact free energy 
           (Eq.~\eqref{eq:FSeries}, full dots) along with the three corresponding coefficients obtained within the
           weak anchoring limit (Eq.~\eqref{eq:FWC}, open circles) for the coupling
           constant $c=0.02$ (upper panel) and $c=0.22$ (lower panel).}
   \label{fig:FCoeffs1}
\end{figure}

Figure~\ref{fig:FCoeffs1} shows the absolute value of the coefficients $a_n$ as functions of $n$ 
(full dots) for $c=0.02$ (upper panel) and $c=0.22$ (lower panel). 
In addition, Fig.~\ref{fig:FCoeffs1} also displays the corresponding 
coefficients obtained within the weak anchoring limit (see Eq.~\eqref{eq:FWC} and the open circles): 
$a_0/2 = \pi c - 4c^2/3 + \mathcal{O}(c^3)$, $a_1 = -\pi c + \mathcal{O}(c^3)$,
and $a_2 = 4c^2/3 + \mathcal{O}(c^3)$.
For small values of the coupling constant $c$ (see the case $c=0.02$ in the upper panel)
the agreement between the weak anchoring coefficients and the exact ones is
excellent.
On the other hand, for large coupling constants (see $c=0.22$ in the lower
panel) one finds (i) that modes appear with comparatively large amplitudes $|a_n|$, $n\geq 3$, which signals that the exact data cannot be strictly described within the
weak anchoring limit given by Eq.~\eqref{eq:FWC}, and (ii) that the exact coefficients $|a_n|$, $n\in\{0,1,2\}$, are not perfectly reproduced by those calculated
within the weak anchoring limit according to Eq.~\eqref{eq:FWC}. On a logarithmic scale these features are not conspicuous. However, they are much more apparent on a linear scale (not
shown here). 
Both of these observations suggest that upon increasing the coupling constant
$c$, Eq.~\eqref{eq:FWC} has to be modified in a way that (i) higher-order 
terms proportional to $\cos(2n\theta),n\geq3$, occur and (ii) terms of order
$c^3$ or higher modify the coefficients $a_n$, $n=0,1,2$.

\begin{figure}[t!]
   \centering
   \includegraphics[scale=1.0]{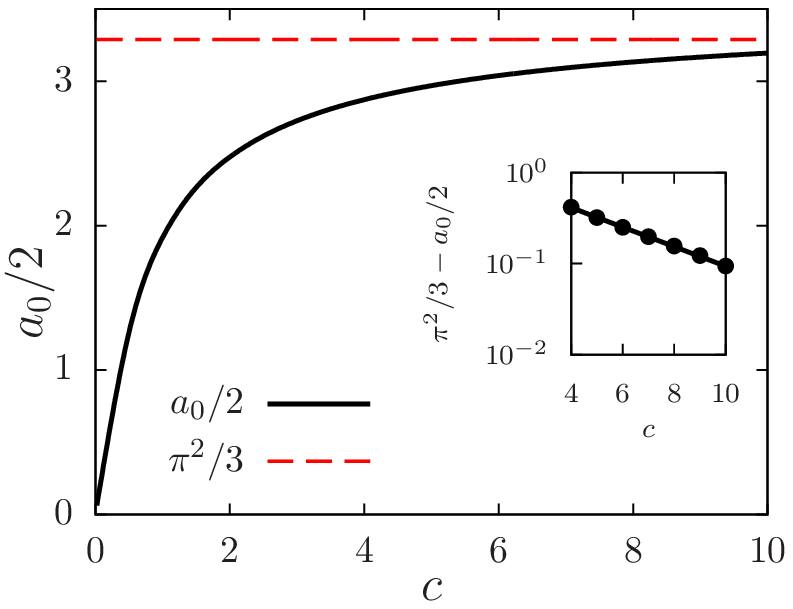}
   \caption{Coefficient $a_0$ of the leading term in the expansion in Eq.~\eqref{eq:FSeries} of
           the free energy as function of the 
           coupling constant $c$. 
           The limit $a_0(c)/2 \to \pi^2/3$ for $c\to\infty$ is indicated by
           the red horizontal dashed line. 
           The inset displays a semi-logarithmic plot of the deviation of $a_0/2$ from its asymptotic value, which is very well fitted by an exponential 
           function $\lambda\exp(-\chi c)$ (solid line), where $\lambda=1.11 \pm 0.02$ and 
           $\chi=0.247\pm 0.003$.}
   \label{fig:Cross}
\end{figure}

As a byproduct of the Fourier analysis presented above, the very strong 
anchoring limit can be reconsidered.
Figure \ref{fig:Cross} shows the coefficient $a_0/2$ appearing in Eq.~\eqref{eq:FSeries} as 
function of $c$. 
As expected, for very strong couplings this curve approaches the value
$\pi^2/3$, which is the coefficient $a_0/2$ appearing in the Fourier expansion of 
Eq.~\eqref{eq:Hayes}. 
This limiting value is attained exponentially (see the inset of
Fig.~\ref{fig:Cross}).

\begin{figure}[t]
   \centering
   \includegraphics[scale=1.0]{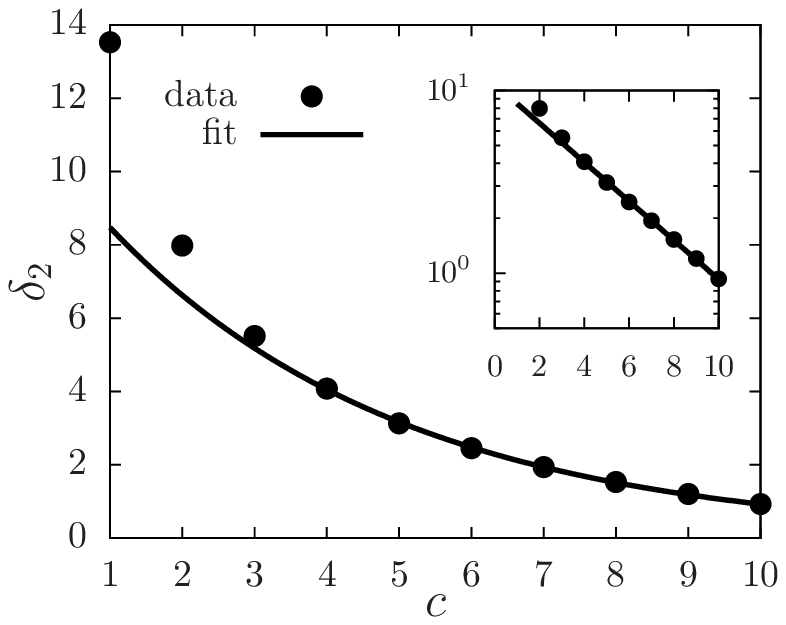}
   \caption{Quadratic norm $\delta_2$ (see Eq.~\eqref{eq:Criterion2}) as 
           function of the coupling constant $c$ (full dots). 
           The solid line is a fit of the data in the interval $c\in[4,10]$ by
           an exponential function $\kappa\exp(-\nu c)$; $\kappa=10.84\pm 0.16$ and $\nu=0.246\pm 0.003$. 
           The inset provides the same information but in terms of a semi-logarithmic plot.}
   \label{fig:Criterion2}
\end{figure}

In order to obtain an estimate of $c_{\text{strong}}$ such that the infinite 
anchoring limit in Eq.~\eqref{eq:Hayes} is reliable for $c>c_\text{strong}$, one
can use a criterion similar to the one in Eq.~\eqref{eq:Criterion1}, based on the 
quadratic norm
\begin{equation}
  \label{eq:Criterion2}
  \delta_2(c):=
  \sqrt{\int_0^{\pi/2}\dd\theta\left|\frac{F^\text{IA}}{KR}-
                                     \frac{F}{KR}\right|^2},
\end{equation}
where $F^\text{IA}$ (Eq.~\eqref{eq:Hayes}) is the free energy within the infinite anchoring limit. 
Figure \ref{fig:Criterion2} shows $\delta_2$ as a function of $c$. 
The tail of the data in the interval $c\in[4,10]$ can be fitted by an
exponential function $\kappa\exp(-\nu c)$ with $\kappa=10.84 \pm 0.16$, and $\nu=0.246 \pm 
0.003$ (see the inset of Fig.~\ref{fig:Criterion2}). 
Thus, $\delta_2<\varepsilon=10^{-2}$ leads to
\begin{equation}
  c_{\text{strong}}=
  -\frac{1}{\nu}\ln\frac{\varepsilon}{\kappa} \approx 
  -\frac{1}{0.246}\ln\frac{\varepsilon}{10.84} \approx 
  28.4 \approx 
  28.
\end{equation}

Accordingly, the interval $c\in[0,\infty)$ of coupling constants provides three regimes: (i) the weak coupling regime $c\in[0,c_{\text{weak}}]$ in which
the free energy of a disc-like colloid in an NLC is very well described by 
Eq.~\eqref{eq:FWC}, (ii) the strong coupling regime $c\in[c_{\text{strong}},
\infty)$ in which the free energy is independent of $c$ and has the form of 
Eq.~\eqref{eq:Hayes}, and (iii) the intermediate coupling regime $c\in
[c_{\text{weak}},c_{\text{strong}}]$ in which the crossover between both previous
limits takes place.


\subsection{Free energy functional in quadratic approximation}

\begin{figure}[b]
   \centering
   \includegraphics[scale=1.0]{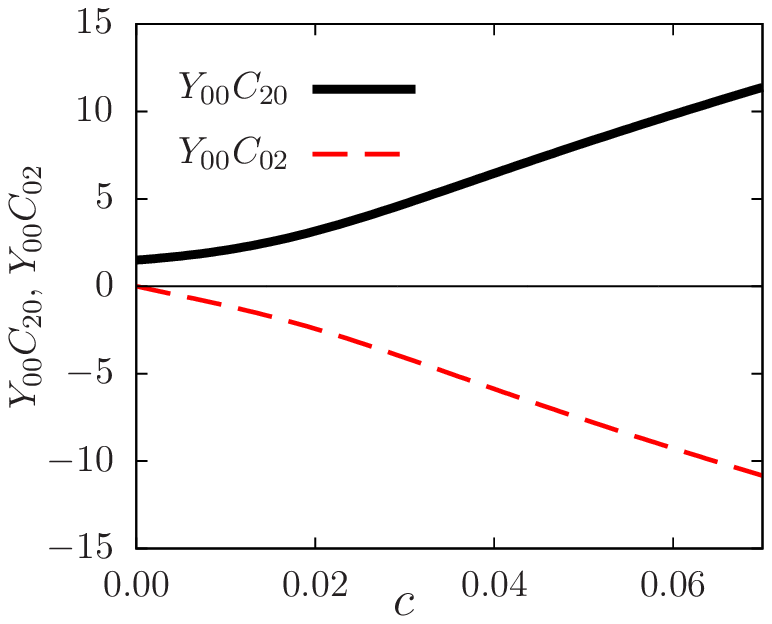}
   \caption{Coefficients $Y_{00}C_{20}$ (Eqs.~\eqref{eq:Y00} and \eqref{eq:C20}) and $Y_{00}C_{02}$ (Eqs.~\eqref{eq:Y00} and \eqref{eq:C02}) in Eq.~\eqref{eq:FinalF2} as
           functions of the coupling constant $c$.}
   \label{fig:Coeffs}
\end{figure}

Disregarding the constant, $\mathbf{M}$- and $\mathbf{n}$-independent term 
$\propto C_{00}$ in the integrand on the right-hand side of Eq.~\eqref{eq:FinalF},
one infers the following expression for the free energy density of a fluid of
magnetic discs suspended in an NLC:
\begin{align}
   \label{eq:FinalF2}
   f(\mathbf{M},\mathbf{n})
   &= k_\text{B}T\rho_\text{iso}\bigg(
   Y_{00}C_{20}\left|\frac{\mathbf{M}}{m\rho_\text{iso}}\right|^2 + \\
   &\ \phantom{= k_\text{B}T\rho_\text{iso}\Big(}
   Y_{00}C_{02}\left(\frac{\mathbf{M}}{m\rho_\text{iso}}\cdot\mathbf{n}
   \right)^2\bigg)
   - \mathbf{M}\cdot\mathbf{B}.
   \notag
\end{align}
The dependence of the coefficients $Y_{00}C_{20}$ and $Y_{00}C_{02}$ (see Eqs.~\eqref{eq:Y00}, \eqref{eq:C20}, and \eqref{eq:C02}) on the coupling 
strength $c$ is shown in Fig.~\ref{fig:Coeffs}.
In contrast, in Ref.~\cite{2013_Mertelj} the following
\textit{ph}enomenological form of the free energy density has been proposed:
\begin{equation}
  \label{eq:Mertelj}
  f_\text{ph}(\mathbf{M},\mathbf{n}) =
  \frac{a}{2}|\mathbf{M}|^2 + \frac{b}{4}|\mathbf{M}|^4
  - \frac{1}{2}\gamma\mu_0(\mathbf{M}\cdot\mathbf{n})^2
  - \mathbf{M}\cdot\mathbf{B},
\end{equation}
where the first two terms on the right-hand side are part of the Landau
expansion describing the interaction between magnetic dipoles, the third term
represents the coupling between the nematic order and the magnetization, and
the last term is the interaction of magnetic dipoles with an external magnetic
field.

The comparison between Eqs.~\eqref{eq:FinalF2} and \eqref{eq:Mertelj} leads to the
following conclusions: 
(i) By identifying the terms proportional to $|\mathbf{M}|^2$ in both 
expressions one infers the positive coefficient $a=2k_\text{B}TY_{00}C_{20}/
(m^2\rho_\text{iso})>0$. 
(ii) Since $a>0$, the term proportional to $|\mathbf{M}|^4$ is unnecessary in the
phenomenological expression in Eq.~\eqref{eq:Mertelj} and its absence in 
Eq.~\eqref{eq:FinalF2} is without consequences.
(iii) In agreement with physical intuition the coefficient $\gamma$ introduced
in Ref.~\cite{2013_Mertelj} is positive and its dependence on the coupling
strength $c$ is given by
\begin{align}
  \gamma
  &=
  -2k_\text{B}T\frac{1}{\rho_\text{iso}}\frac{1}{\mu_0m^2}Y_{00}C_{02}  
  \notag\\
  &=\frac{k_\text{B}T}{2}\frac{1}{\rho_\text{iso}}\frac{1}{\mu_0m^2}Y_{00}
  \frac{Y_{12}/Y_{10}}{Y_{10}+Y_{12}}.
  \label{eq:gamma}
\end{align}

\begin{figure}[t!]
   \centering
   \includegraphics[scale=1.0]{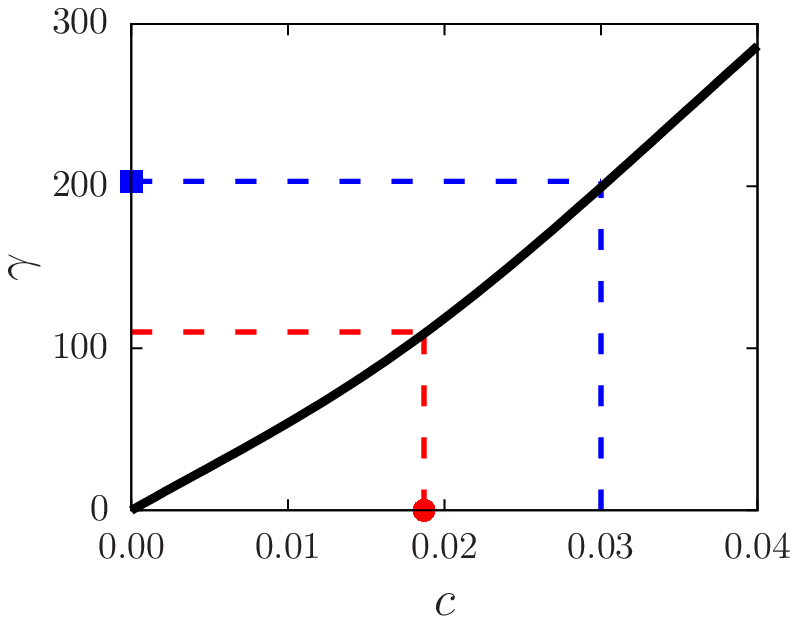}
   \caption{Dependence of the phenomenological coupling coefficient $\gamma$ 
           introduced in Ref.~\cite{2013_Mertelj} (see Eq.~\eqref{eq:Mertelj})
           on the microscopic coupling constant $c$ as it follows from 
           Eq.~\eqref{eq:gamma}. 
           The red circle denotes the value of $c\simeq 0.019$, which corresponds to the 
           estimate of $\gamma\simeq 110$ given in Ref.~\cite{2014_Mertelj}. 
           The blue square denotes the value of $\gamma\simeq 203$, which corresponds to the
           estimate of $c\simeq 0.03$ based on the parameters used in Ref.~\cite{2014_Mertelj}.
           Under ideal circumstances, the blue and the red dashed lines coincide.}
   \label{fig:Gamma}
\end{figure}

Figure \ref{fig:Gamma} shows $\gamma$ as function of the coupling constant $c$
(see Eq.~\eqref{eq:gamma}) for the experimentally relevant parameters 
$\rho_\text{iso}=1.5 \times 10^{19}\,\mathrm{m^{-3}}$ and $m \approx 3 \times
10^{-18}\,\mathrm{Am^2}$ taken from Ref.~\cite{2014_Mertelj}.
The order of magnitude of the theoretical result (black line) is in agreement
with the values estimated from the experiment ($\gamma_\text{exp}\approx 110$,
see Ref.~\cite{2014_Mertelj}). 
Figure~\ref{fig:Gamma} shows that (at least in the regime of weak anchoring) 
$\gamma$ increases monotonically with $c$. 
Based on the values given in Ref.~\cite{2014_Mertelj} one can, on the one hand,
estimate $c$ from $\gamma$ (red dot and dashed lines) and, on the other hand,
one can estimate $\gamma$ from $c$ (blue square and dashed lines). 
Since values of $c$ and $\gamma$ given in Ref.~\cite{2014_Mertelj} belong to 
one and the same system the red and blue dashed lines in Fig.~\ref{fig:Gamma}
should coincide. 
However, this is not quite the case.
The discrepancy may arise due to the fact that in all calculations the mean value of the particle
size has been used assuming that the discs are monodisperse in size, whereas in the experiment the size distribution of the colloids 
has a finite width.
Moreover, elastic interactions between the discs (generated by the nematic director field $\mathbf{n}$), which have been entirely 
neglected in the present study, might play a role for the properties of the actual system.

Knowing the explicit dependence $\gamma(c)$ offers the possibility to estimate
the anchoring energy $W$ by performing an experiment similar to the one 
described in Ref.~\cite{2014_Mertelj}: Using Fig.~\ref{fig:Gamma},
from an estimate of $\gamma$ one obtains the corresponding value of $c$, which, knowing the
mean size $R$ of the platelets and the elastic constant $K$ of the NLC, renders the value of $W$.

\begin{figure}[t!]
   \centering
   \includegraphics[scale=1.0]{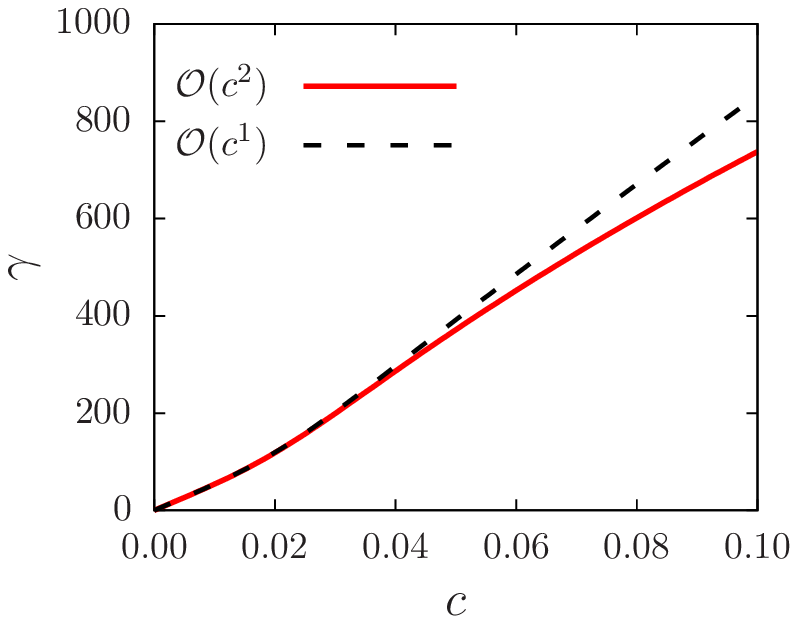}
   \caption{Comparison of the values of $\gamma$ calculated by using in 
           Eq.~\eqref{eq:VNLC} only the first-order term in $c$ (black dashed line, $\mathcal{O}(c^1)$)
           or both the first- and the second-order term in $c$ (red solid line, $\mathcal{O}(c^2)$). 
           The larger $c$, the more does the first-order theory overestimate the
           value of $\gamma$.}
   \label{fig:Orders}
\end{figure}

In Fig.~\ref{fig:Orders} two expressions of $\gamma$ as function of $c$ are
compared: one, for which in Eq.~\eqref{eq:VNLC} only the term of first order in $c$ (black dashed
line, $\mathcal{O}(c^1)$) and one, for which terms up to second order in $c$ (red solid line, 
$\mathcal{O}(c^2)$) are retained. (Note that the black dashed line is a \textit{nonlinear} function of $c$, even if one has kept only the linear contribution $\mathcal{O}(c^1)$ in $V_\text{NLC}$, because $\gamma$ depends nonlinearly on $V_\text{NLC}$.)
The difference between the two expressions suggests that if one cuts off the expansion in $c$ at too low order, the
resulting phenomenological coupling constant $\gamma$ is overestimated if $c\gtrsim 0.04$. The value of $c$ inferred from the experiment (see Ref.~\cite{2014_Mertelj}) lies below the value $c=0.04$, i.e., in a range within which the two curves in Fig.~\ref{fig:Orders} de facto coincide. Therefore, considering only the term $\propto c^1$ in Eqs.~\eqref{eq:FWC} or \eqref{eq:VNLC} does not change the outcome of the present effective theory in the context of the experimental parameters used in Ref.~\cite{2014_Mertelj}. However, this assessment requires an analysis up to higher orders in $c$, as carried out in the present study.


\section{\label{sec:Discus}Discussion}

Inspired by an expression introduced in Ref.~\cite{2013_Mertelj}, the present
study derives a Landau-like free energy density of ferronematics in terms of the
magnetization $\mathbf{M}$ and the nematic director $\mathbf{n}$.
The derivation starts from a density functional theory (DFT) which
describes colloids suspended in a nematic liquid crystal (NLC).
The coupling between the colloids and the NLC is modeled in terms of a 
one-particle potential (see Eq.~\eqref{eq:VNLC}), which in the DFT framework
plays the role of an external field. It depends on the orientation of the
colloid and on the local nematic director field $\mathbf{n}(\mathbf{r})$.
Motivated by the high dilution of the colloidal suspensions under consideration,
a direct colloid-colloid interaction is neglected. 
Accordingly, the theory can be formulated in terms of a relatively simple local 
density functional. 
The one-particle potential in Eq.~\eqref{eq:VNLC} is derived from the
perturbation expansion (Eq.~\eqref{eq:FWC}) of the free energy in terms of the small 
parameter $c$, which represents the strength of the coupling of the NLC to the
surface of a single colloidal particle (see Eq.~\eqref{eq:C}). 
In the present study the expansion of the free energy in terms of powers of $c$, together with the corresponding 
analytical expression for the nematic director profile around a colloidal particle, is determined. The term $\propto c^1$ in the expansion derived here is equivalent to the expression derived elsewhere (see Ref.~\cite{2014_Tasinkevych}) for the different case of arbitrarily thin rods with tangential coupling. 
Using numerical methods, the range of values of the coupling constant $c$ is
estimated, within which the weak coupling limit (Eq.~\eqref{eq:FWC}) is accurate.
It is shown that the next-order term is proportional to 
$\big(\sin (3\theta)\big)^2$ with $\theta$ introduced in Fig.~\ref{fig:Colloid}.

In the next step, the expression in Eq.~\eqref{eq:VNLC} for the one-particle potential
is used to establish the density functional in Eq.~\eqref{eq:MicroF}
of noninteracting discs subjected to an external field. 
Four possible kinds of pair interactions are neglected: 
(i) the direct dipole-dipole interaction due to the presence of magnetic moments; (ii) the steric repulsive interaction; (iii) the van der Waals interaction; and 
(iv) the effective elastic interaction induced by the NLC. 
On one hand, the dipole-dipole and the van der Waals interactions are negligibly small compared to the thermal energy
$k_\text{B}T$ for the mean distances between the disc centers as given in 
Ref.~\cite{2014_Mertelj}. 
Moreover, the steric interaction is disregarded due to its short-ranged character and therefore due to the low impact onto mesoscopic properties of a very dilute colloidal solution.
On the other hand, the effective elastic interaction might be important even for dilute 
solutions because the effective elastic interaction for two discs, which are both
inclined with respect to the far-field director, is described by a long-ranged 
Coulomb-like pair potential \cite{2007_Pergamenshchik}. 
Here, we have neglected it nevertheless in order to keep the theory analytically tractable.

\begin{figure}[t!]
   \centering
   \includegraphics[scale=1.0]{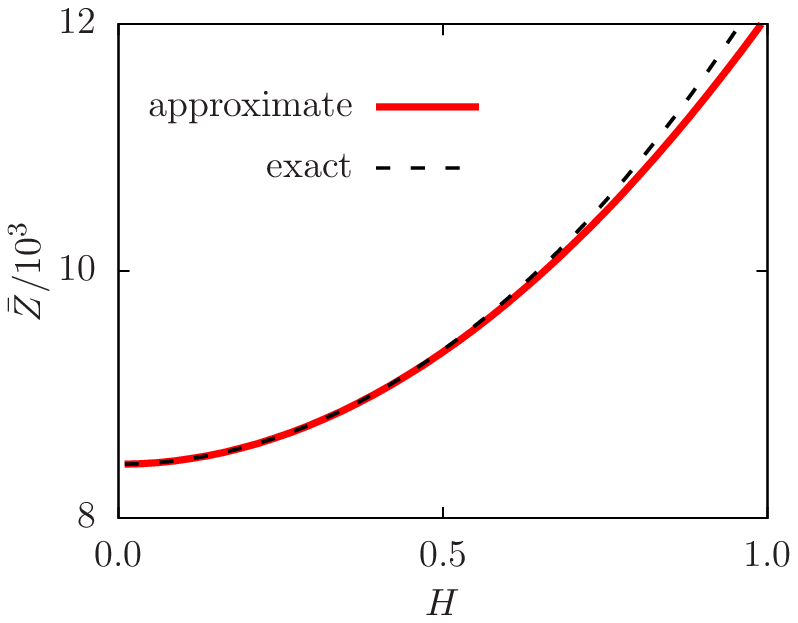}
   \caption{Comparison of the exact generating function $\bar{Z}(H,u)$ given by 
           Eq.~\eqref{eq:ExactZ} (black dashed line) and its quadratic 
           approximation provided in Eq.~\eqref{eq:ApproxZ} (red solid line) for 
           $c=0.05$ and $u=1$.}
   \label{fig:HMax}
\end{figure}

Deriving the free energy density in terms of powers of $|\mathbf{M}|$ and of the
scalar product $\mathbf{M}\cdot\mathbf{n}$ appears to be out of analytic reach because the integral in Eq.~\eqref{eq:ExactZ} and thus $Z(\mathbf{h})$ cannot be calculated analytically. 
Hence, in Eq.~\eqref{eq:ApproxZ} the approximation of $Z(\mathbf{h}) = 
\bar{Z}(H,u)$ by a second-degree polynomial in terms of the variables 
$H:=m|\mathbf{h}|$ and $u:=\mathbf{n}\cdot\mathbf{h}/|\mathbf{h}|$ is 
introduced. 
The quality of this quadratic approximation is high for small values  of $H$ and
it becomes poorer upon increasing $H$. 
Note, however, that in the present study large values of $H$ are not needed.
Indeed, whereas one can show that within the present model one has $|\mathbf{M}|\to\infty$ for $H\to\infty$,
this limit is not realized in the context of the experimental situation under
consideration, because the colloids carry a finite magnetic dipole moment 
$m$ and their suspension in the NLC is given by a finite number density 
$\rho_\text{iso}$, so that $|\mathbf{M}(\mathbf{r})|\leq m\rho_\text{iso}$ at any
point $\mathbf{r}$ (for reason of simplicity assuming only very small segregation effects). 
Therefore one should not consider the whole range of $H$ but only up to a
certain value $H_\text{max}$, defined such that $|\mathbf{M}|/(m\rho_\text{iso})\leq
T(H_\text{max})/Y_{00}:=1$ holds (see the definition of $T$ in the paragraph below Eq.~\eqref{eq:Fugacity}, and see Eq.~\eqref{eq:Fugacity}). 
According to the second line of Eq.~\eqref{eq:tT}, $H_\text{max}$ is a function
of $u\in[-1,1]$, i.e., of the cosine of the angle between the effective 
magnetic field $\mathbf{h}$ and the nematic director $\mathbf{n}$ of the NLC (see below Eq.~\eqref{eq:Fugacity}). 
Figure \ref{fig:HMax} shows a comparison of the exact generating function 
$\bar{Z}(H,u)$ (black dashed line) with the one calculated within the quadratic 
approximation (red solid line) in the interval $H\in[0,H_\text{max}(u)]$ for
$u=1$, i.e., for the case that the magnetization is parallel to the nematic
director.
It turns out that the difference between the exact generating function and its
quadratic approximation increases with $H$ but that it remains below 10\% for the
whole range of physically reasonable values of $H$. 
Thus the quadratic approximation in Eq.~\eqref{eq:ApproxZ} provides a reasonable,
at least qualitatively correct description of the free energy density of ferronematics.

The free energy-density obtained in the present study (Eq.~\eqref{eq:FinalF2}) and
the one in Eq.~\eqref{eq:Mertelj} proposed in Ref.~\cite{2013_Mertelj} share
two main features:
(i) In the absence of the coupling between the magnetization $\mathbf{M}$ and the nematic
director $\mathbf{n}$, the suspension is in the paramagnetic phase, and 
(ii) the ferromagnetic properties are generated by a coupling of the magnetization
and the nematic director via a term proportional to 
$(\mathbf{M}\cdot\mathbf{n})^2$.
The comparison of the coefficients multiplying $(\mathbf{M}\cdot\mathbf{n})^2$ 
allows one to obtain an expression for the phenomenological coupling parameter
$\gamma$ (Eq.~\eqref{eq:Mertelj}) in terms of the microscopic coupling constant $c$ (Eq.~\eqref{eq:gamma}). 
A slight inconsistency between the estimates for $\gamma$ and for the anchoring
energy $W$ from Ref.~\cite{2014_Mertelj} is found (Fig.~\ref{fig:Gamma}). The reasons for this 
discrepancy encompass both the simplifications used in the present study
(such as arbitrarily thin discs, monodisperse disc size, neglect of elastic pair
interactions) as well as those applied in the theoretical model used in 
Ref.~\cite{2014_Mertelj}, e.g., the simplified form of the coupling of the director field to
the colloid surface.

\appendix

\section{\label{sec:A1}Boundary problems for $\alpha^{(0)}$ and $\alpha^{(1)}$}

Solving the boundary problem in Eq.~\eqref{eq:BP} analytically is difficult due
to the nonlinearity of the boundary condition at the disc surface.
However, the expansion in Eq.~\eqref{eq:Expan} of the scalar field $\alpha$ in
terms of powers of the coupling constant $c$ generates a set of boundary 
problems corresponding to $\alpha^{(0)}, \alpha^{(1)}, \dots$, which are much
simpler.
In the following the boundary problems for $\alpha^{(0)}$ and $\alpha^{(1)}$
are solved, and the free energy in Eq.~\eqref{eq:AFunc} is determined by inserting the expansion into Eq.~\eqref{eq:Expan} up to terms $n\leq 1$, i.e., with the equilibrium expressions for $\alpha^{(0)}$ and $\alpha^{(1)}$.


\subsection{Zeroth order in $c$}

The boundary problem corresponding to $\alpha^{(0)}$ is posed as
\begin{equation}
  \label{eq:BP0}
  \begin{cases}
      \boldsymbol{\nabla}^2\alpha^{(0)}(\mathbf{r})=0
    & \text{, $\mathbf{r}\in\mathcal{V}$}\\
      \boldsymbol{\nabla}\alpha^{(0)}(\mathbf{s})\cdot\boldsymbol{\nu}
      (\mathbf{s})=0
    & \text{, at disc surface}\\
      \alpha^{(0)}(\mathbf{r})\simeq\theta
    & \text{, $|\mathbf{r}|\gg R$}.
  \end{cases}
\end{equation}

From Eq.~\eqref{eq:Expan} one infers $\alpha=\alpha^{(0)}$ for $c=0$ which
corresponds to the limit of a decoupling of the liquid crystal and the colloid.
Therefore, in this limit the nematic director $\mathbf{n}(\mathbf{r})$ is
not distorted by the presence of the colloidal disc, i.e., physical intuition 
leads to the uniform scalar field 
\begin{equation}
  \label{eq:A0}
  \alpha^{(0)}(\mathbf{r}) = \theta.
\end{equation}
It can be readily verified that this is indeed the solution of Eq.~\eqref{eq:BP0}.


\subsection{First order in $c$}

Using Eq.~\eqref{eq:A0}, the boundary problem corresponding to $\alpha^{(1)}$
is
\begin{equation}
    \label{eq:BP1}
    \begin{cases}
        \boldsymbol{\nabla}^2\alpha^{(1)}(\mathbf{r})=0
      & \text{, $\mathbf{r}\in\mathcal{V}$}\\
        \dps\boldsymbol{\nabla}\alpha^{(1)}(\mathbf{s})\cdot\boldsymbol{\nu}
        (\mathbf{s})=\frac{\sin(2\alpha^{(0)}(\mathbf{s}))}{R}
      & \\[5pt]
        \dps\phantom{\dps\boldsymbol{\nabla}\alpha^{(1)}(\mathbf{s})\cdot
        \boldsymbol{\nu}(\mathbf{s})}
        =\frac{\sin(2\theta)}{R}
      & \text{, at disc surface}\\
        \alpha^{(1)}(\mathbf{r})\simeq0
      & \text{, $|\mathbf{r}|\gg R$}.
    \end{cases}
\end{equation}

By identifying $\alpha^{(1)}(\mathbf{r})$ with an electrostatic potential 
$\lambda\varphi(\mathbf{r})$, where $\lambda$ is a constant with the dimension of
an inverse voltage, one can map Eq.~\eqref{eq:BP1} into the problem of finding
the electrostatic potential $\varphi$ of an arbitrarily thin, uniformly charged
disc with radius $R$ and surface charge density $\sigma = 
-\varepsilon_0\sin(2\theta)/(\lambda R)$.
This is given by \cite{2011_Ciftja}
\begin{equation}
  \varphi(\rho,z) =
  \frac{\sigma R}{\varepsilon_0}\int_0^{\infty}\frac{dk}{k}
  J_0(k\rho)J_1(kR)\exp(-k|z|),
\end{equation}
where $\rho:=\sqrt{x^2+y^2}$ is the distance from the $z$-axis and where $J_i$
denotes the Bessel function of order $i$. 
With the necessary replacements one obtains the solution of 
Eq.~\eqref{eq:BP1} in the form
\begin{equation}
  \label{eq:A1}
  \alpha^{(1)}(\rho,z) =
  -\sin(2\theta)\int_0^{\infty}\frac{\dd k}{k}\,\,J_0(k\rho)J_1(kR)\exp(-k|z|).
\end{equation}


\subsection{Free energy}

By using Eq.~\eqref{eq:A0} in Eq.~\eqref{eq:Expan}, the expansion of the scalar
field $\alpha$ in terms of powers of the coupling constant $c$ is given by
\begin{equation}
  \alpha(\mathbf{r}) = \theta + c\alpha^{(1)}(\mathbf{r}) + \mathcal{O}(c^2).
\end{equation}
Inserting this expression into the free energy functional in Eq.~\eqref{eq:AFunc}
one obtains
\begin{align}
  \label{eq:FrankOseen}
  \frac{F[\alpha]}{KR} \!
  &= \frac{c^2}{2R}\int_{\mathcal{V}}\!\!\dd^3 r\,
  \left[\boldsymbol{\nabla}\alpha^{(1)}(\mathbf{r})\right]^2 \!+\! 
  \frac{c}{R^2}\!\int_{\partial\mathcal{V}}\!\!\!\!\!\dd^2\! s\,
  [\sin(\alpha(\mathbf{s}))]^2 \notag\\
  &\ \ \ + \mathcal{O}(c^3),
\end{align}
where
\begin{align}
  [\sin(\alpha(\mathbf{s}))]^2
  &= [\sin(\theta + c\alpha^{(1)}(\mathbf{s}) + \mathcal{O}(c^2))]^2 
  \notag\\
  &= [\sin\theta + c\cos\theta\,\alpha^{(1)}(\mathbf{s}) + \mathcal{O}(c^2)]^2
  \label{eq:sin2}\\
  &= (\sin\theta)^2+2c\sin\theta\cos\theta\,\alpha^{(1)}(\mathbf{s})
     + \mathcal{O}(c^2).
  \notag
\end{align}

In order to calculate the volume integral in Eq.~\eqref{eq:FrankOseen} one can
use Green's first identity
\begin{align}
  &\int_\mathcal{V}\!\!\dd^3 r\,\,
  \left[\alpha^{(1)}(\mathbf{r})\bs{\nabla}^2\alpha^{(1)}(\mathbf{r})+
  \bs{\nabla}\alpha^{(1)}(\mathbf{r})\cdot\bs{\nabla}\alpha^{(1)}(\mathbf{r})
  \right]
  \notag\\
  =
  &\int_{\partial\mathcal{V}}\!\!\dd^2 s\,\, 
  \alpha^{(1)}(\mathbf{s})\bs{\nabla}\alpha^{(1)}(\mathbf{s})\cdot
  \bs{\kappa}(\mathbf{s})
  \label{eq:GreenUsed}
\end{align}
where $\bs{\kappa}(\mathbf{s})$ is the outer normal at the point $\mathbf{s}\in
\partial\mathcal{V}$. 
The first term in the volume integral in Eq.~\eqref{eq:GreenUsed} vanishes due
to the first line of Eq.~\eqref{eq:BP1}.
Moreover,
\begin{widetext}
\begin{align}
  \label{eq:Green}
  \int_{\partial\mathcal{V}}\!\!\dd^2 s\,\, 
  \alpha^{(1)}(\mathbf{s})\bs{\nabla}\alpha^{(1)}(\mathbf{s})\cdot
  \bs{\kappa}(\mathbf{s})
  =
  \int_{|\mathbf{s}|=\,\text{const}\gg R}\hspace{-4em}\dd^2 s\,\,
  \alpha^{(1)}(\mathbf{s})\bs{\nabla}\alpha^{(1)}(\mathbf{s})\cdot
  \bs{\kappa}(\mathbf{s}) -
  \int_{\text{disc}}\!\!\dd^2 s\,\,
  \alpha^{(1)}(\mathbf{s})\bs{\nabla}\alpha^{(1)}(\mathbf{s})\cdot
  \bs{\nu}(\mathbf{s}),
\end{align}
where $\bs{\nu}(\mathbf{s}) = -\kappa(\mathbf{s})$ is the normal of the disc
surface.
The first integral on the right-hand side of Eq.~\eqref{eq:Green} vanishes
due to the third line of Eq.~\eqref{eq:BP1}. 
This leads to
\begin{equation}
  \frac{c^2}{2R}\int_{\mathcal{V}}\!\!\dd^3\! r\,
  \left[\boldsymbol{\nabla}\alpha^{(1)}(\mathbf{r})\right]^2 
  = - \frac{c^2}{2R}\int_{\text{disc}}\!\!\dd^2 s\,\,
  \alpha^{(1)}(\mathbf{s})\bs{\nabla}\alpha^{(1)}(\mathbf{s})\cdot
  \bs{\nu}(\mathbf{s}).
  \label{eq:volumeterm}
\end{equation}
Finally, by using Eqs.~\eqref{eq:BP1}, \eqref{eq:sin2}, and 
\eqref{eq:volumeterm}, Eq.~\eqref{eq:FrankOseen} turns into
\begin{align}
  \frac{F[\alpha]}{KR} \!
  &= \int_{\text{disc}}\!\!\dd^2 s\,\,
  \left[  \frac{c}{R^2}(\sin\theta)^2 
        + \frac{2c^2}{R^2}\sin\theta\cos\theta\,\alpha^{(1)}(\mathbf{s})
        - \frac{c^2}{2R}\alpha^{(1)}(\mathbf{s})
          \bs{\nabla}\alpha^{(1)}(\mathbf{s})\cdot\bs{\nu}(\mathbf{s})
  \right] + \mathcal{O}(c^3)\nonumber\\
  &= 2\pi c(\sin\theta)^2 + \frac{c^2}{2R^2}\sin(2\theta)
     \int_{\text{disc}}\!\!\dd^2 s\,\,\alpha^{(1)}(\mathbf{s}).
\end{align}
Inserting Eq.~\eqref{eq:A1} with $z=0$ into the last integral one obtains
\begin{align}
  \int_{\text{disc}}\!\!\dd^2 s\,\,\alpha^{(1)}(\mathbf{s})
  &= 
  2\int_0^R\!\!\dd \rho\,\,\rho\int_0^{2\pi}\!\!\dd\varphi\,\,
  (-\sin (2\theta))\int_0^\infty\!\!\frac{\dd k}{k}\,\,J_0(k\rho)J_1(kR)\nonumber\\
  &=
  -4\pi R\sin(2\theta)\int_0^\infty\!\!\frac{\dd k}{k^2}\,\,
  \left(J_1(kR)\right)^2\nonumber\\
  &=
  -\frac{16}{3}R^2\sin(2\theta),
  \label{eq:2Sides}
\end{align}
(the prefactor of 2 in front of the integral in the first term on the right-hand side of Eq.~\eqref{eq:2Sides} accounts for the two faces of the disc surface) so that
\begin{align}
  \frac{F}{KR}
  &= 2\pi c(\sin\theta)^2 - \frac{8}{3}c^2(\sin(2\theta))^2 + 
  \mathcal{O}(c^3)
  \notag\\
  \label{eq:OneParticleFromA}
  &= \text{const} - (2\pi c + \frac{32}{3}c^2)(\mathbf{n}_{0}\cdot\bs{\omega})^2
  + \frac{32}{3}c^2(\mathbf{n}_{0}\cdot\bs{\omega})^4+\mathcal{O}(c^3),
\end{align}
which, upon ignoring the irrelevant constant term, leads to Eq.~\eqref{eq:FWC}.


\section{\label{sec:A2}Quadratic approximation of the generating function 
        $Z(\mathbf{h})$}

In the following we provide a detailed derivation of Eq.~\eqref{eq:FinalF} within the
quadratic approximation (see Eq.~\eqref{eq:ApproxZ}) of the generating function introduced in
Eq.~\eqref{eq:ExactZ}.

The generating function $Z(\mathbf{h})$ in 
Eq.~\eqref{eq:ExactZ} can be rewritten as
\begin{equation}
  \label{eq:ZSimp}
  Z(\mathbf{h}) =
  \int\!\!\dd^2\omega\,\,
  f(\mathbf{n}(\mathbf{r})\cdot\bs{\omega})\exp(\mathbf{h}\cdot m\bs{\omega})
\end{equation}
with $f(x) := \exp(A_1x^2 + A_2x^4)$.

As a first step, we show that $Z(\mathbf{h})$ is an even function 
$\bar{Z}(H,u)$ of both $H:=m|\mathbf{h}|$ and $u:=\mathbf{n}\cdot\mathbf{h}/
|\mathbf{h}|$. To this end we consider an appropriate coordinate system such that the $z$-axis points along the local director field $\mathbf{n}(\mathbf{r})$ and the $x$-axis is chosen in an arbitrary direction in the plane perpendicular to $\mathbf{n}(\mathbf{r})$. (Note the difference in the meaning of $\theta$ and $\alpha$ between Eq.~\eqref{eq:Coord} and Fig.~\ref{fig:Colloid}.)
\begin{equation}
  \mathbf{n}=\begin{bmatrix}
    0\\
    0\\
    1
  \end{bmatrix},
  \quad
  \bs{\omega}=\begin{bmatrix}
    \sin\theta\cos\varphi\\
    \sin\theta\sin\varphi\\
    \cos\theta
  \end{bmatrix},
  \quad
  \mathbf{h}=\frac{H}{m}\begin{bmatrix}
    \sin\alpha\cos\beta\\
    \sin\alpha\sin\beta\\
    \cos\alpha
  \end{bmatrix}=:\frac{H}{m}\mathbf{v}
  \label{eq:Coord}
\end{equation}
so that $|\mathbf{v}|=1$ and $u=\mathbf{n}\cdot\mathbf{v}=\cos\alpha$.
With this choice Eq.~\eqref{eq:ZSimp} takes the form
\begin{align}
  Z(\mathbf{h})
  &=
  \int_0^\pi\!\!\dd\theta\,\,\sin\theta f(\cos\theta)
  \int_0^{2\pi}\!\!\!\!\dd\varphi\exp\left[H(\sin\alpha\sin\theta
  \cos(\varphi-\beta)+\cos\alpha\cos\theta)\right]
  \notag\\
  &=
  2\pi\int_{-1}^1\!\!\dd x\,\, f(x)I_0(H\sqrt{1-u^2}\sqrt{1-x^2})\exp(Hux)
  \label{eq:ZForNum}\\
  &=:\bar{Z}(H,u), \notag
\end{align}
where $I_0$ is a modified Bessel function of order $0$ (see
Ref.~\cite{Gradshteyn1}, Eq.~(8.431.3)).
Since $f(x)$ is an even function of $x$, one can infer from
Eq.~\eqref{eq:ZForNum} that $\bar{Z}(H,u)$ is an even function of both $H$ and $u$.
\end{widetext}

Using Eq.~\eqref{eq:MZ} one can express the magnetization $\mathbf{M}$ in 
terms of $H$ and $u$:
\begin{equation}
  \frac{\mathbf{M}}{\zeta} =
  \frac{\partial Z}{\partial \mathbf{h}} =
  \frac{\partial H}{\partial \mathbf{h}}\frac{\partial \bar{Z}}{\partial H} +
  \frac{\partial u}{\partial \mathbf{h}}\frac{\partial \bar{Z}}{\partial u}
\end{equation}
with
\begin{align}
  &\frac{\partial H}{\partial h_i} =
  \frac{\partial}{\partial h_i}m|\mathbf{h}| =
  m\frac{h_i}{|\mathbf{h}|}
  \notag\\
  \Rightarrow\  
  &\frac{\partial H}{\partial \mathbf{h}} = 
  m\frac{\mathbf{h}}{|\mathbf{h}|} = m\mathbf{v}
\end{align}
and
\begin{align}
  &\frac{\partial u}{\partial h_i} =
  \frac{\partial}{\partial h_i}\ 
  \frac{\mathbf{n}\cdot\mathbf{h}}{|\mathbf{h}|} =
  \frac{n_i}{|\mathbf{h}|} - 
  (\mathbf{n}\cdot\mathbf{h})\frac{h_i}{|\mathbf{h}|^3}
  \notag\\
  \Rightarrow\ 
  &\frac{\partial u}{\partial \mathbf{h}} =
  \frac{1}{|\mathbf{h}|}\left(\mathbf{n} -
  \frac{\mathbf{n}\cdot\mathbf{h}}{|\mathbf{h}|}\ 
  \frac{\mathbf{h}}{|\mathbf{h}|}\right) =
  \frac{m}{H}(\mathbf{n}-u\mathbf{v}).
\end{align}

In the next step, we consider the quantities $T=|\mathbf{M}|/(m\zeta)$ and 
$t=\mathbf{n}\cdot\mathbf{M}/(m\zeta)$ which are related to $H$ and $u$ via
\begin{align}
  t 
  &= \frac{\mathbf{n}\cdot\mathbf{M}}{m\zeta}
  = u\frac{\partial\bar{Z}}{\partial H}+
     \frac{1-u^2}{H}\frac{\partial\bar{Z}}{\partial u}
  \notag\\
  T^2 
  &= \left(\frac{\mathbf{M}}{m\zeta}\right)^2
  = \left(\frac{\partial\bar{Z}}{\partial H}\right)^2 +
     \frac{1-u^2}{H^2}\left(\frac{\partial\bar{Z}}{\partial u}\right)^2.
  \label{eq:Tandt}
\end{align}

Since an analytical expression for the integral in Eq.~\eqref{eq:ExactZ}
is not available, it is rewritten as a series in powers of 
$|\mathbf{h}|$:
\begin{widetext}
\begin{align}
  Z(\mathbf{h})
  &=
  \int\!\!\dd^2\omega\,\, \exp(A_1(\mathbf{n}\cdot\bs{\omega})^2 +
                               A_2(\mathbf{n}\cdot\bs{\omega})^4)
                          \exp(\mathbf{h}\cdot m\bs{\omega}) 
  \notag\\
  &=
  \int\!\!\dd^2\omega\,\, \exp(A_1(\mathbf{n}\cdot\bs{\omega})^2 +
                               A_2(\mathbf{n}\cdot\bs{\omega})^4)
  \sum_{k=0}^\infty \frac{1}{k!}(\mathbf{h}\cdot m\bs{\omega})^k 
  \notag\\
  &= 
  \sum_{k=0}^\infty \frac{1}{(2k)!} \int\!\!\dd^2\omega\,\, 
  \exp(A_1(\mathbf{n}\cdot\bs{\omega})^2 +
       A_2(\mathbf{n}\cdot\bs{\omega})^4)
  (\mathbf{h}\cdot m\bs{\omega})^{2k} 
  \notag\\
  &=
  \sum_{k=0}^\infty Z_k(H,u).
  \label{eq:Zseries}
\end{align}
We note that $Z_k=0$ for $k$ odd. Since here the ultimate goal is to derive Eq.~\eqref{eq:FinalF}, expressions of
$H$ and $u$ in terms of $T$ and $t$ are required, which are obtained by 
inverting the map $(H,u)\to(T,t)$ in Eq.~\eqref{eq:Tandt}.
However, an inversion of Eq.~\eqref{eq:Tandt} in closed form is feasible only
when the series in Eq.~\eqref{eq:Zseries} is restricted to sufficiently low
orders.
In the following only the terms $Z_k(H,u)$ with $k \leq 1$ are considered.
The term $Z_{k=0}(H,u)$ in Eq.~\eqref{eq:Zseries} is given by
\begin{align}
  Z_0(H,u) 
  &=
  \int\!\!\dd^2\omega\,\, \exp(A_1(\mathbf{n}\cdot\bs{\omega})^2 +
                               A_2(\mathbf{n}\cdot\bs{\omega})^4)
  \notag\\
  &= 
  \int_0^{2\pi}\!\!\!\!\dd\varphi\int_0^\pi\!\!\dd\theta\,\,\sin\theta 
  \exp(A_1\cos^2\theta+A_2\cos^4\theta)
  \notag\\
  &=
  2\pi\int_{-1}^1\!\!\dd x\,\, \exp(A_1x^2 + A_2x^4)
  \notag\\
  &=:
  2\pi \mathcal{I}_0(A_1,A_2)
  \notag\\
  &=:
  Y_{00},
  \label{eq:Y00}
\end{align}
whereas the term $Z_{k=1}(H,u)$ in Eq.~\eqref{eq:Zseries} is given by
\begin{align}
    Z_1(H,u)
    &=
    \frac{1}{2}\int\!\!\dd^2\omega\,\, 
    \exp(A_1(\mathbf{n}\cdot\bs{\omega})^2 +
         A_2(\mathbf{n}\cdot\bs{\omega})^4)
    (\mathbf{h}\cdot m\bs{\omega})^2
    \notag\\
    &=
    \frac{H^2}{2}\int\!\!\dd^2\omega\,\, 
    \exp(A_1(\mathbf{n}\cdot\bs{\omega})^2 + 
         A_2(\mathbf{n}\cdot\bs{\omega})^4)
    (\sin\alpha\sin\theta\cos(\varphi-\beta) + \cos\alpha\cos\theta)^2
    \notag\\
    &=
    \frac{H^2}{2}\int\!\!\dd^2\omega\,\, 
    \exp(A_1(\cos\theta)^2 + A_2(\cos\theta)^4)
    \Big[(\sin\alpha\sin\theta\cos(\varphi-\beta))^2 + 
    \notag\\
    &\ \phantom{=\frac{H^2}{2}\int\!\!\dd^2\omega\,\,
              \exp(A_1(\cos\theta)^2 + A_2(\cos\theta)^4)\Big[}
     2\sin\alpha\cos\alpha\sin\theta\cos\theta\cos(\varphi-\beta) +
    \notag\\
    &\ \phantom{=\frac{H^2}{2}\int\!\!\dd^2\omega\,\,
              \exp(A_1(\cos\theta)^2 + A_2(\cos\theta)^4)\Big[}
     (\cos\alpha\cos\theta)^2\Big]
    \notag\\
    &=
    \frac{H^2}{2}\int_{-1}^1\!\!\dd x\,\, 
    \exp(A_1 x^2 + A_2 x^4)
    \left[\pi(1-x^2)(\sin\alpha)^2 + 2\pi x^2(\cos\alpha)^2\right]
    \notag\\
    &=
    \pi\frac{H^2}{2}\left[
    (1-u^2) \int_{-1}^1\!\!\dd x\,\, \exp(A_1 x^2+A_2 x^4) +
    (3u^2-1)\int_{-1}^1\!\!\dd x\,\, x^2\exp(A_1 x^2+A_2 x^4)
    \right]
    \notag\\
    &=:
    \pi\frac{H^2}{2}
    \left[(1-u^2)\mathcal{I}_0(A_1,A_2) +
          (3u^2-1)\mathcal{I}_1(A_1,A_2)\right]
    \notag\\
    &=:
    Y_{10}H^2 + Y_{12}H^2u^2
\end{align}
\end{widetext}
with
\begin{align}
  Y_{10} :=
  \frac{\pi}{2}\left(\mathcal{I}_0-\mathcal{I}_1\right),\quad
  Y_{12} :=
  \frac{\pi}{2}\left(3\mathcal{I}_1-\mathcal{I}_0\right).
\end{align}
This leads to the ``quadratic'' approximation
\begin{align}
  \bar{Z}(H,u) 
  &\approx Z_0(H,u) + Z_1(H,u) \notag\\
  &= Y_{00} + Y_{10}H^2 + Y_{12}H^2u^2.
  \label{eq:quadraticZ}
\end{align}

Inserting Eq.~\eqref{eq:quadraticZ} into Eq.~\eqref{eq:Tandt} one obtains
\begin{equation}
  \begin{cases}
    t^2&=4(Y_{10}+Y_{12})^2H^2u^2 \\
    T^2&=4Y_{10}^2H^2+(8Y_{10}Y_{12}+4Y_{12}^2)H^2u^2.
  \end{cases}
  \label{eq:tTHu}
\end{equation}
which leads to
\begin{equation}
  \begin{cases}
    H^2u^2
    &= 
    \frac{t^2}{4(Y_{10}+Y_{12})^2}
    \\
    H^2
    &=
    \frac{1}{4Y_{10}^2}\left(T^2-
      \frac{t^2(2Y_{10}Y_{12}+Y_{12}^2)}{(Y_{10}+Y_{12})^2}\right).
  \end{cases}
\end{equation}
Finally, that part of the integrand in 
Eq.~\eqref{eq:NewNewMicroF}, which depends on $Z$, is
\begin{align}
  \mathbf{h}\cdot\frac{\partial Z}{\partial \mathbf{h}}-Z
  &= 
  H\frac{\partial \bar{Z}}{\partial H}-\bar{Z}
  \notag\\
  &= C_{00}+C_{20}T^2+C_{02}t^2,
  \label{eq:hdZdhmZ}
\end{align}
where
\begin{align}
  \label{eq:C00}
  C_{00}&:=-Y_{00}\\
  \label{eq:C20}
  C_{20}&:=\frac{1}{4Y_{10}}\\
  \label{eq:C02}
  C_{02}&:=-\frac{Y_{12}/Y_{10}}{4(Y_{10}+Y_{12})},
\end{align}
which, upon insertion into Eq.~\eqref{eq:NewNewMicroF}, leads to 
Eq.~\eqref{eq:OmegaTt}.
Equation~\eqref{eq:FinalF} follows from expressing $T$ and $t$ in terms
of $|\mathbf{M}|$ and $\mathbf{M}\cdot\mathbf{n}$.

As expected, the quadratic approximation becomes poorer the larger $H$ is. 
However, it turns out to be a reasonable approximation within the physically relevant range of $H$ (see Sec.~\ref{sec:Discus}). 
In contrast, if in Eq.~\eqref{eq:Zseries} one keeps terms with $k>1$, $\big(T(H,u)\big)^2$ and
$\big(t(H,u)\big)^2$ in Eq.~\eqref{eq:tTHu} are polynomials of at least degree 2 in $H^2$
and $H^2u^2$.
In this case $H^2$ and $H^2u^2$ are not polynomials in $T^2$ and $t^2$, which
implies that Eq.~\eqref{eq:hdZdhmZ}, and therefore the integrand given in
Eq.~\eqref{eq:FinalF2} for Eq.~\eqref{eq:FinalF}, is not represented by a 
polynomial and thus cannot be compared with the expression in Eq.~\eqref{eq:Mertelj}.



\begin{thebibliography}{00}
\bibitem{footnote2}
  $V_\text{NLC}$ can be expected to be an even function of the particle normal $\bs{\nu}$ due to the inversion symmetry of the nematic phase (see Subsec.~\ref{sec:Theory1}). This allows one to formulate $V_\text{NLC}$ in terms of the orientation of the magnetic dipole $\bs{\omega}=\pm\bs{\nu}$.
\bibitem{footnote3}
  The nematic director field $\mathbf{n}(\mathbf{r})$ at position $\mathbf{r}$ minimizes the Frank-Oseen functional in Eq.~\eqref{eq:FOFunc} which fulfills the boundary condition $\mathbf{n}(\mathbf{r})=\mathbf{n}_0$ far away from the colloidal disc. The coupling, according to Eq.~\eqref{eq:FS}, of the NLC to the disc surface leads to a distortion of $\mathbf{n}(\mathbf{r})$ in the direction $e_z$ of the disc normal. Hence, in order to avoid disadvantageous splay contibutions, $\mathbf{n}(\mathbf{r})$ at any position $\mathbf{r}$ has to be parallel to the space spanned by $\mathbf{n}_0$ and $e_z$ (see Eq.~\eqref{eq:NAnsatz}).
\bibitem{footnote4}
  Having $\mathbf{M}=0$ everywhere in the sample implies that the density profile $\bar{\rho}$ in Eq.~\eqref{eq:rhobar} is an even function of $\bs{\omega}$ (i.e., there are as many particles with orientation $\bs{\omega}$ as there are particles with orientation $-\bs{\omega}$) which is only possible if $\lambda(\mathbf{r},[0,\mathbf{n}])=0$, provided that $\mathbf{B}=0$.
\bibitem{footnote5}
  Obviously, such a criterion can be chosen in various ways. 
  Moreover, even in the context of the present choice, there is still some degree of
  freedom in choosing $\varepsilon$.
\end{thebibliography}
\end{document}